\documentclass{pasa}%

\title[Why do some cores remain starless ?]{Why do some cores remain starless ?}
\author[Anathpindika, S]{Anathpindika, S$^1,^2$ \thanks{sumed$\_$k@yahoo.co.in}\\
\affil{$^1$Department of Physics, Indian Institute of Technology, Kharagpur 721302, $^2$ Institute for Astronomy \& Astrophysics(IAAT), University of T$\ddot{\mathrm{u}}$bingen, 10 Auf Der MorgenStelle, T$\ddot{\mathrm{u}}$bingen, Germany}}%
\jid{PASA}
\doi{10.1017/pas.\the\year.xxx}
\jyear{\the\year}

\begin{document}%
\begin{abstract}
Prestellar cores, by definition, are gravitationally bound but starless pockets of dense gas. Physical conditions that could render a core starless(in the local Universe) is the subject of investigation in this work. To this end we studied the evolution of four starless cores, {\small B68, L694-2, L1517B, L1689}, and {\small L1521F}, a {\small VeLLO}. The density profile of a typical core extracted from an earlier simulation developed to study core-formation in a molecular cloud was used for the purpose. We demonstrate - (\textbf{i}) cores contracted in quasistatic manner over a timescale on the order of $\sim 10^{5}$ years. Those that remained starless did briefly acquire a centrally  concentrated density configuration that mimicked the density profile of a unstable Bonnor Ebert sphere before rebounding,  (\textbf{ii}) three of our test cores viz. L694-2, L1689-SMM16 and L1521F remained starless despite becoming thermally super-critical. On the contrary B68 and L1517B remained sub-critical; L1521F collapsed to become a VeLLO only when gas-cooling was enhanced by increasing the size of dust-grains. This result is robust, for other cores viz. {\small B68, L694-2, L1517B} and \small{L1689} that previously remained starless could also be similarly induced to collapse.  Our principle conclusions are : (\textbf{a}) acquiring the thermally super-critical state does not ensure that a core will necessarily become protostellar, (\textbf{b}) potentially star-forming cores (the {\small VeLLO} {\small L1521F} here), could be experiencing coagulation of dust-grains that must enhance the gas-dust coupling and in turn lower the gas temperature, thereby assisting collapse. This hypothesis appears to have some observational support, and (\textbf{c}) depending on its dynamic state at any given epoch, a core could appear to be pressure-confined, gravitationally/virially bound, suggesting that gravitational/virial boundedness of a core is insufficient to ensure  it will form stars, though it is crucial for gas in a contracting core to cool efficiently so it can collapse further to become protostellar. Gas temperature in these purely hydrodynamic calculations was calculated by explicitly solving the respective equations of thermal-balance for gas and dust; the attenuation factor for the interstellar radiation-field, $\chi$, which in literature has been well constrained to $\sim 10^{-4}$ for putative star-forming clumps and cores was adopted in these calculations.
\end{abstract}
\begin{keywords}
ISM : individual objects : B68, L694-2, L1517B, L1689, L1517B  -- Starless cores -- Star-formation
\end{keywords}
\maketitle%

\section{INTRODUCTION }
\label{sec:intro}
It is well-known that a number of cores do in fact remain starless. Key physical characteristics about these cores to have emerged from various detailed observational surveys are - \textbf{(a)} their longevity or in other words, the fact that some cores remain starless for a period significantly longer than others do (e.g. Ward-Thompson \emph{et al.} 2007), and \textbf{(b)} that cores typically are not pockets of isothermal molecular gas, but are colder close to their respective centres. Also, gas within cores is usually sub-sonic or transonic at best (e.g. Evans \emph{et al.} 2001, Crapsi \emph{et al.} 2004; 2007, di Francesco \emph{et al.} 2007, Schnee \emph{et al.} 2013). Observational estimates of the typical age of cores derived from detailed studies of their chemical composition and the strength of the magnetic field threading them are on the order of $\sim 10^{6}$ years, which is on the order of a free-fall time for a typical prestellar core (e.g. Beichman \emph{et al.}1986, Kirk \emph{et al.} 2005, Ward-Thompson \emph{et al.}2007; see e.g. Maret \emph{et al.} 2013 for an estimate of the chemical age). Broadly speaking, models attempting to explain the evolution of prestellar cores may be classified into two paradigms : \textbf{(1)} those that propose quasistatic evolution of magnetised cores, and \textbf{(2)} those in which cores form and evolve on a relatively short time-scale in turbulent clouds, the so-called paradigm of rapid star-formation. \\ \\
Depending on the strength of magnetic field threading a core, it may be further classified as magnetically sub-critical(mass to flux ratio less than unity), or super-critical(mass to flux ratio greater than unity). Sub-critical cores collapse on the ambipolar diffusion timescale (e.g. Shu \emph{et al.} 1987, Mouschovias 1991), while those that are super-critical, in the absence of turbulence, collapse essentially on the free-fall timescale (e.g. Nakano 1998). This model, however, encounters a number of difficulties : \textbf{(i)} the sub-critical nature of cores is inconsistent with the findings of a detailed study of a handful of isolated cores that show these cores are threaded by only a relatively weak magnetic field (e.g. Kirk \emph{et al.} 2006), \textbf{(ii)} the ambipolar diffusion timescale, typically on the order of $\sim 10^{7}$ yrs - the lifetime of a molecular cloud (e.g. Nakano 1998, Ciolek \& Basu 2000), is much higher than the estimates of typical core lifetime, and \textbf{(iii)} it is also inconsistent with the paradigm of rapid star-formation, reinforced by the observations of several nearby star-forming regions where stellar populations are found to have ages significantly smaller than the crossing-time for their respective host cloud (e.g. Hartmann \emph{et al.} 2001, Elmegreen 2000).  \\ \\
In the dynamic picture of star-formation, on the other hand,
supersonic turbulence, perhaps moderated by a weak magnetic field, largely regulates formation of prestellar cores as they condense out via amplification of local perturbations in the density field of molecular clouds. Also, since turbulence decays rapidly, over less than a free-fall time, cores supported by turbulence could possibly explain various features of starless cores (e.g. Mac Low \emph{et al.}1998, Ballesteros-Paredes \emph{et al.} 2003; 2006). Cores, according to this model, are in approximate (thermal) pressure-equilibrium with the ambient medium and have a density distribution that reflects the profile of a pressure-confined {\small BES}. One of the difficulties with this model of core-formation is the transient nature of resulting cores on account of being merely in  hydrostatic equilibrium (e.g. Mac Low \& Klessen 2004), which is inconsistent with typical properties of starless cores. Furthermore, simulations such as those by Walch \emph{et al.}(2010), for instance, demonstrate that turbulence can only delay collapse in gravitationally super-critical cores, but cannot arrest it.  Evidently a solution to the problem of starless nature of cores must be sought beyond these tested ideas such that their physical properties noted above can be reconciled. \\ \\
Evolution of cores has been studied analytically by a number of authors over the last four decades. Some of the earliest models suggested by e.g., Bodenheimer \& Sweigert (1968), Larson (1969) and Penston(1969) were criticised later by Shu (1977) on grounds of idealised(artificial) initial conditions. The Shu model that envisaged a test core modelled as a unstable Bonnor-Ebert sphere({\small BES}) which would then collapse in the \emph{inside-out} manner was itself later deemed physically unrealisable, for it is difficult to reconcile the assembly of a core that is critically stable at best, and unstable at worst (Whitworth \& Summers 1985; hereafter WS). WS demonstrated analytically that a pressure-confined {\small BES} could indeed be driven to collapse in \emph{outside-in} fashion by a compressional wave triggered by raising sufficiently the magnitude of the externally confining pressure, $P_{ext}$. This was later demonstrated numerically by  Hennebelle \emph{et al.}(2003). Physically, this would be the case in regions of active star-formation where energy and momentum is dumped into the ambient medium by ongoing episodes of star-formation. Similar arguments were presented by G{\' o}mez \emph{et al.}(2007) who demonstrated that a sufficiently strong inwardly propagating compressional wave could induce a {\small BES} to collapse. More recently, Anathpindika \& Di Francesco (2013) showed that a core modelled as a non-isothermal {\small BES} and with sufficiently cold interiors can also collapse in this manner to become protostellar. Other cores in their work where gas close to the core-centre was not sufficiently cold remained starless and in fact, oscillated radially. Similarly, Kaminski \emph{et al.} (2014) demonstrated that {\small BES}s could be similarly induced to collapse by increasing the magnitude of $P_{ext}$, effected by raising the density of the confining medium. They also showed that {\small BES}s enveloped by a hot tenuous medium would likely evolve in quasistatic manner before eventually collapsing. \\ \\
 Models invoking detailed radiative transfer modelling of starless cores include those suggested by Keto \& Caselli (2008, 2010). These authors classified starless cores into two classes viz., thermally sub-critical(central density, $n_{c}\lesssim 10^{5}$ cm$^{-3}$) and super-critical($n_{c}\gtrsim 10^{5}$ cm$^{-3}$) cores. They argue, gas temperature in centrally dense cores is lowered more efficiently via collisional coupling between gas and dust which renders such cores thermally super-critical and therefore, more susceptible to collapse which is likely to proceed in the outside-in manner (e.g. Williams \emph{et al.} 1999, Caselli \emph{et al.} 2002, Schnee \emph{et al.} 2007). On the other hand, thermally sub-critical cores are likely to be relatively stable against self-gravity and may exhibit radial oscillations (e.g. Lada \emph{et al.} 2003, Aguti \emph{et al.}2007). These models usually assume a density distribution to model the temperature profile of a typical starless core. Other relatively simple analytic models such as the Plummer-type suggested by Whitworth \& Ward-Thompson (2001) produces rapid density enhancement on a timescale much shorter than the observational estimates of contraction timescale. \\ \\
In the present work we therefore propose to numerically investigate the evolution of a set of cores whose physical characteristics are well-known. We are particularly interested in studying the conditions under which a core is likely to be rendered unstable against self-gravity. To this end we will self-consistently deduce the temperature profile for a core that remains starless and one that collapses. Also, we will study the temporal excursion of the virial components of our test cores to examine if gravitational and/or virial boundedness is sufficient to ensure that a  prestellar core becomes protostellar. This article is organised as follows : in \S 2 we will discuss the choice of our initial conditions and the numerical scheme employed to investigate the problem. We will present our results and discuss them in respectively \S 3, and \S 4 before concluding in \S 5.
\section{Physical details}
\subsection{The model core : A brief review of the formation of cores and their density profile}
Test cores in the present work were modelled by examining the density profile of a typical core that forms in a molecular cloud. We therefore refer to our recent work (Anathpindika 2015), where the formation of cores had been demonstrated via an interplay between the thermal instability ({\small TI}) and the gravitational instability ({\small GI}), in a purely hydrodynamic realisation of a self-gravitating molecular cloud ({\small MC}).
 For a comprehensive discussion of the subject the interested reader is referred to a recent review by Offner et al. (2014). Lack of adequate computational resources, however, prevented those calculations from continuing further to track the evolution of individual cores. We therefore extracted cores emblematic of those found in typical star-forming regions from an earlier simulation reported in Anathpindika(2015).  Cores could be isolated (e.g. {\small B68}), or be part of an elongated filament-like natal cloud such as the countless number of them detected in the Ophiuchus {\small MC} or the Taurus {\small MC} among many others. \\ \\
The images on panels of Fig. 1 show the two representative cores selected from a simulation reported in Anathpindika (2015), of which, that on the left-hand panel is an isolated core while the one on the right-hand panel belongs to a larger natal filament-like clump. The density profile of each of the two has been shown in Fig. 2. Interestingly, and equally significantly the gas distribution in either core, as is visible from the respective panels of Fig. 2, is very well approximated by the density profile of a stable Bonnor-Ebert sphere({\small BES}) of radius, $\xi_{b}$=3. This is also consistent with the density profiles reported by Teixeira \emph{et al.}(2005) for their sample of cores observed in the Lupus 3 molecular cloud. We use this finding to model our test cores in this work. \\ \\
\begin{figure*}
\vspace{1pc}
\mbox{\includegraphics[angle=270,width=0.5\textwidth]{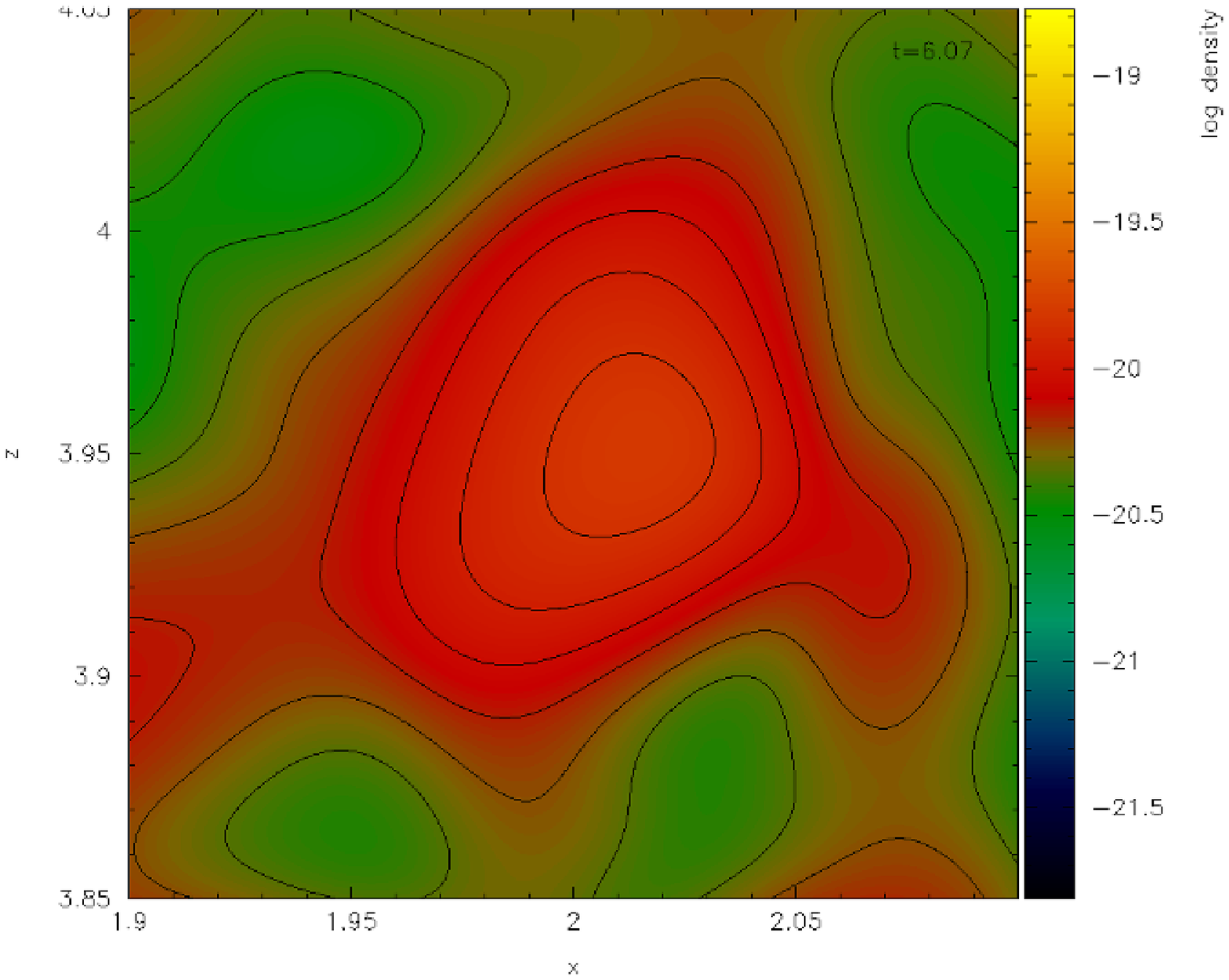}
\includegraphics[angle=270,width=0.5\textwidth]{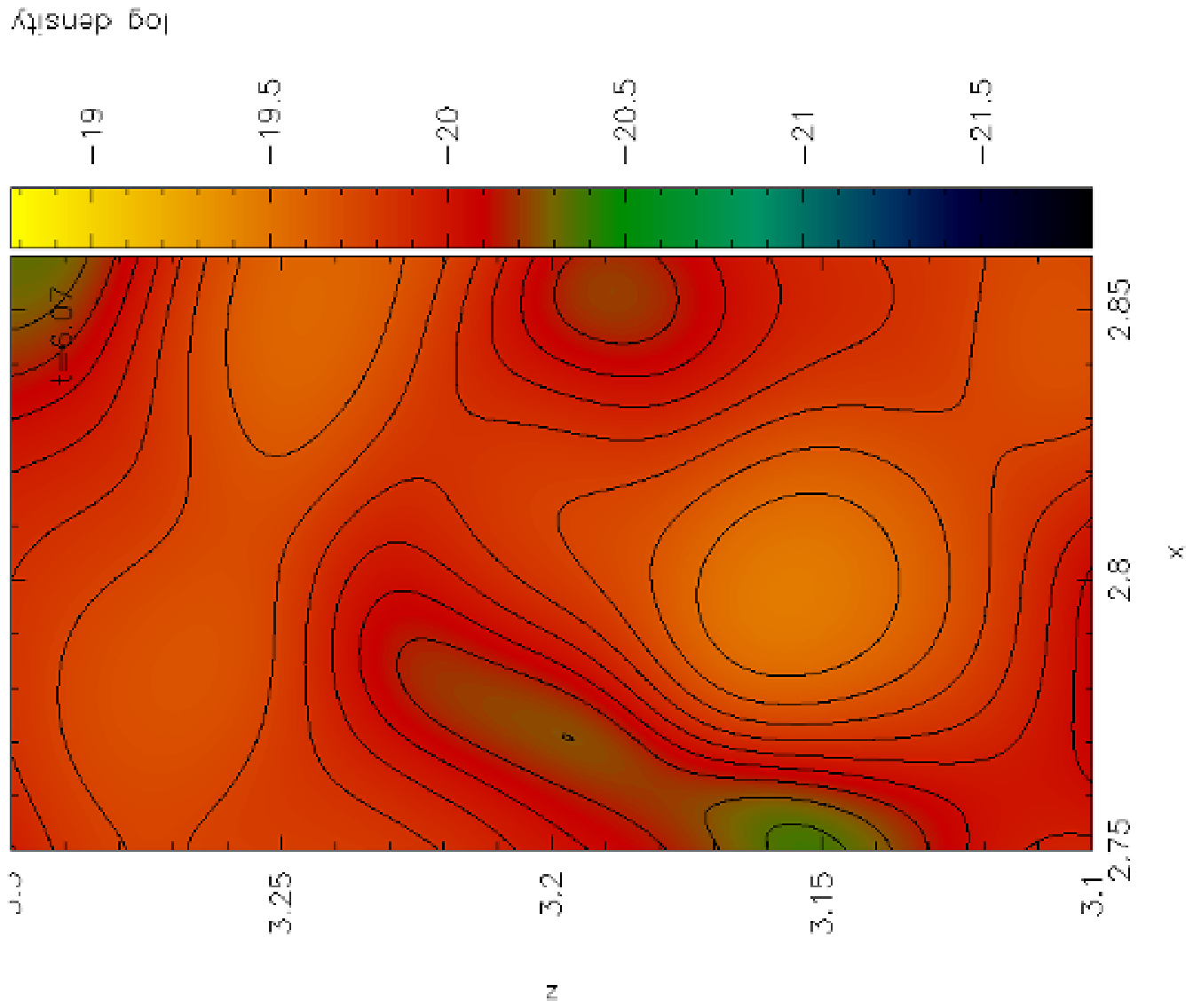}}
\caption{Shown here are rendered density plots from an earlier work (Anathpindika 2015), of a core that formed in a small section of the fragmented cloud. Picture on the left-hand panel is an example of a core in relative isolation, dissociated from any large clump,  while that on the right-hand panel shows a core that belongs to a contiguous, elongated clump in a different region of the same fragmented cloud; time in units of Myr has been marked on the top left-hand corner of each panel.}
\end{figure*}
\begin{figure*}
\vspace{1pc}
\mbox{\includegraphics[angle=270,width=0.5\textwidth]{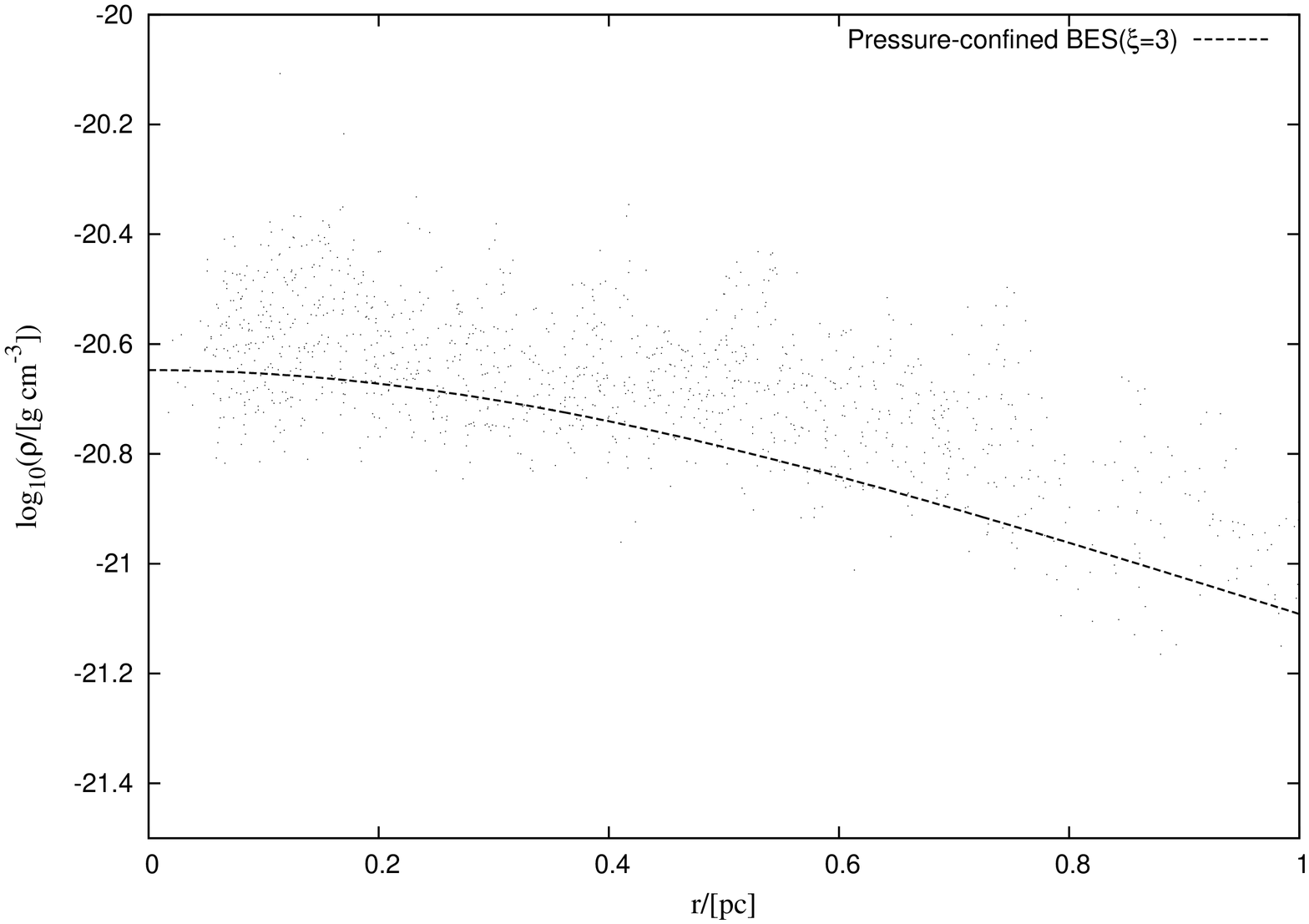}
\includegraphics[angle=270,width=0.5\textwidth]{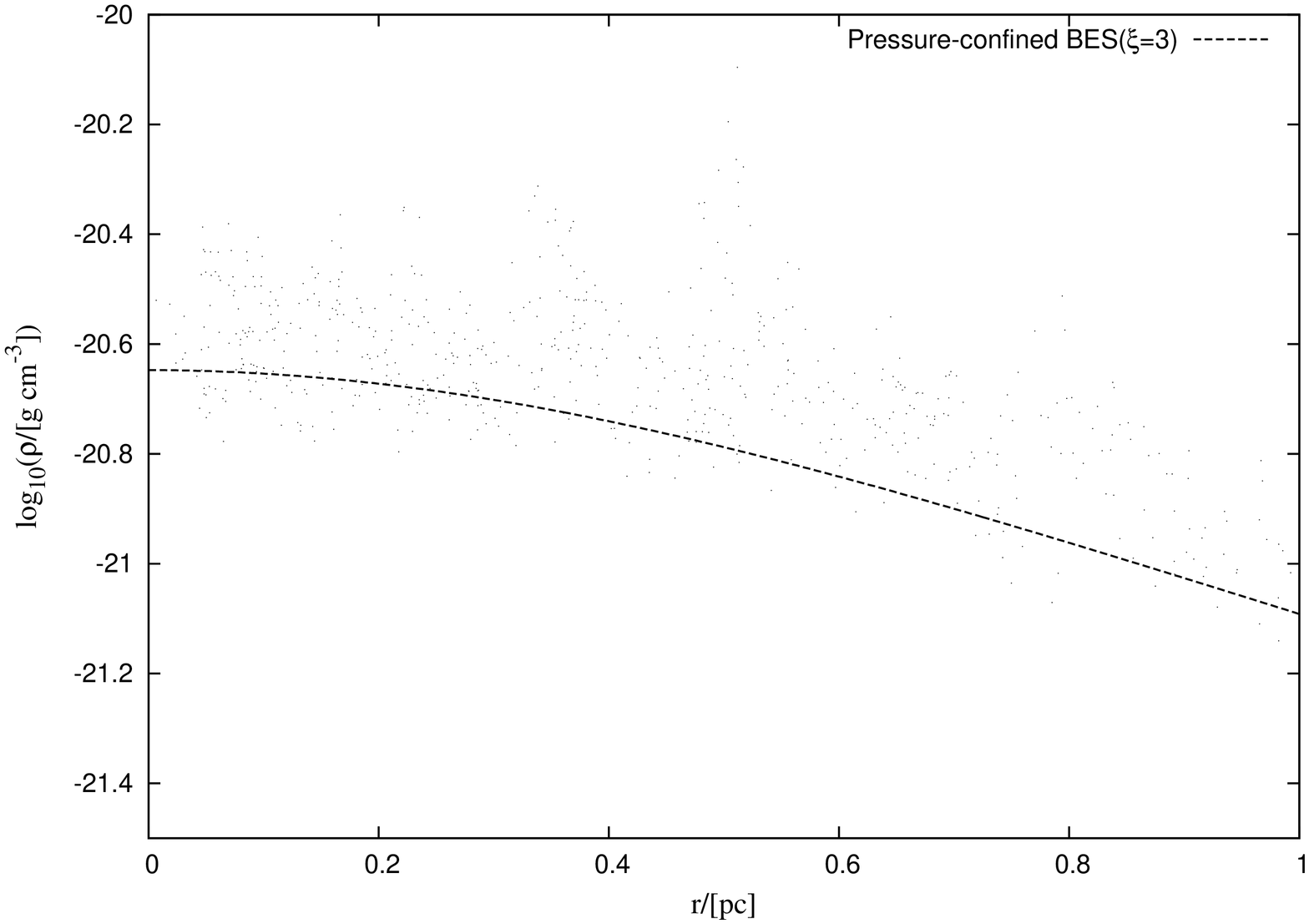}}
\caption{Distribution of gas density in the respective cores shown in Fig. 1 above. Dots here represent density of individual gas particles that assembled these cores and interestingly, the density profile of a stable Bonnor-Ebert sphere having radius, $\xi_{b}$ = 3 (see text below), appears to fit the density distribution of either cores very well.}
\end{figure*}
\begin{figure*}
\vspace{1pc}
\mbox{\includegraphics[angle=270,width=0.5\textwidth]{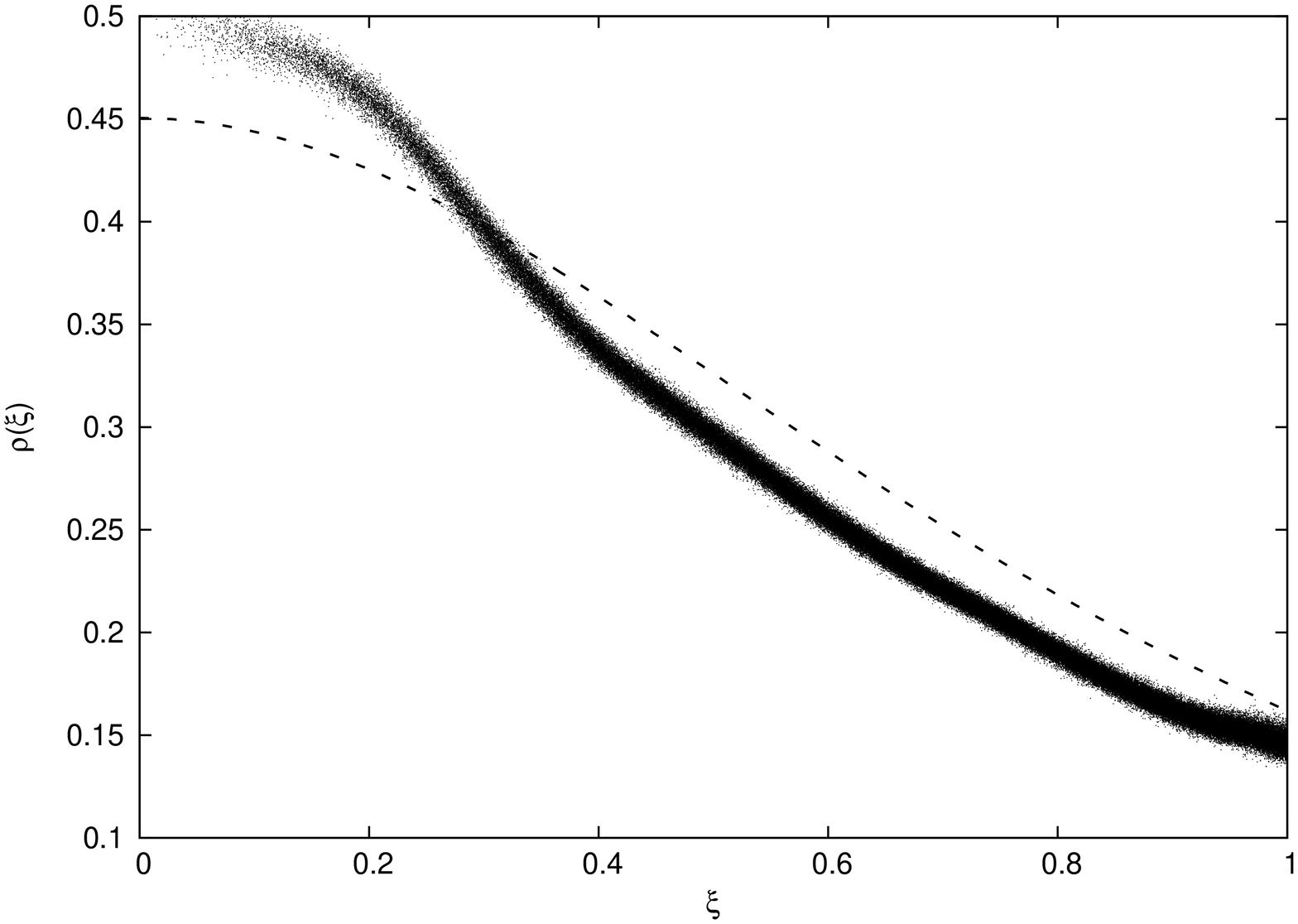}
\includegraphics[angle=270,width=0.5\textwidth]{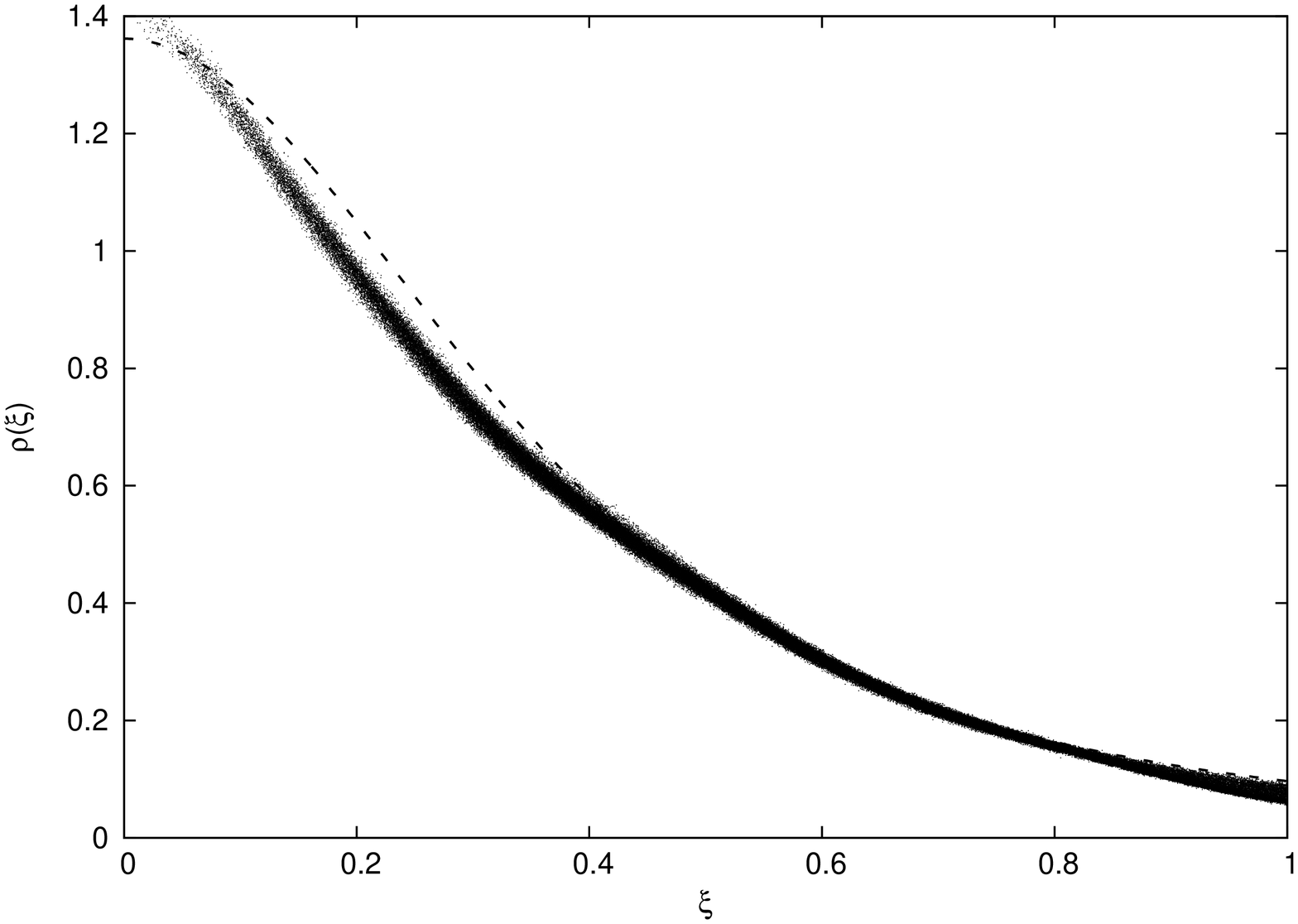}}
\caption{Density profile of the settled stable Bonnor-Ebert sphere ($\xi_{b}$=3), and the critically stable Bonnor-Ebert sphere ($\xi_{b}$=6.45) has been shown respectively, on the left and the right-hand panel. Dots represent the density of gas particles assembled in either polytrope while the analytically expected density distribution given by Eqn. (3) has been shown by a continuous line in these respective plots.}
\end{figure*}
\begin{table*}
 \centering
 \begin{minipage}{140mm}
  \caption{Physical details of test cores modelled in this work; listed at the bottom of the table are the respective references from where physical details of these cores have been obtained.}
  \begin{tabular}{@{}llrrlllll}
  \hline
   Serial & Core &$\frac{M_{core}}{[M_{\odot}]}$ & $\frac{R_{core}}{[pc]}$ &$\frac{T_{gas}}{[K]}$ &$\frac{T_{ext}}{[K]}$ & Q$_{init}$\footnote{Q$=\frac{E_{g}}{E_{p}}$, ratio of gravitational energy with energy due to confinement by external pressure.} & $\frac{P_{ext}(\times 10^{4})}{[K cm^{-3}]}$ & $\frac{t_{ff}}{Myr}$\footnote{Free-fall time for the initial configuration; $t_{ff}=\Big(\frac{3\pi}{32G\bar{\rho}}\Big)^{1/2}, \bar{\rho}$ - average initial gas density in a test core.} \\
   No.    & Name & & & & & &  &\\
   \hline
1($\xi_{b}$=3) & B68\footnote{Alves \emph{et al.}(2001), Bergin \emph{et al.}(2006)}   & 2.1 & 0.15 & 15 & 45& 0.47 & 3.47 & 0.7 \\
2($\xi_{b}$=6.5)& B68 & 2.1 & 0.13  & 15 & 210 & 0.1 & 49.0 & 0.65 \\
3(Uniform density) & B68 & 2.1 & 0.06 & 15 & 45& 1.8 & 3.47 & 0.16 \\
4($\xi_{b}$=3) & L694-2\footnote{Harvey \emph{et al.} (2003), Chitsazzadeh (2014)} & 3.0 & 0.15 & 19 & 55 & 0.5 & 6.05  & 0.56 \\
5($\xi_{b}$=3) & L1517B\footnote{Tafalla\emph{et al.}2004; Kirk \emph{et al.}2006} & 0.4 & 0.05 & 12 & 35 & 0.38 & 1.4 & 0.31\\
6($\xi_{b}$=3) & L1689\footnote{Chitsazzadeh \emph{et al.} (2014)}  & 3.1 & 0.20 & 15 & 45 & 0.51 & 2.6 & 0.88 \\
7($\xi_{b}$=3) & L1521F\footnote{Chitsazzadeh (2014)} & 4.7 & 0.21 & 18 & 50 & 0.61 & 3.25 & 0.78 \\
\hline
   With puffed-up dust grains \\ 
\hline
8($\xi_{b}$=3)(VeLLO) & L1521F & 4.7 & 0.21 & 18 & 50 & 0.61 & 3.25 & 0.78   \\
9 & Same as 1  & & & & & & \\ 
10 &  Same as 4  & & & & & & \\ 
11 & Same as 5  & & & & & & \\ 
12 & Same as 6  & & & & & & \\ 
  \hline
   \end{tabular}
\end{minipage}
\end{table*}
\textbf{The Bonnor-Ebert sphere ({\small BES})}\\
The maximum stable mass of the isothermal (or the Bonnor-Ebert ({\small BE})) sphere of a given radius, $\xi_{b}$, is
\begin{equation}
M_{BE}\equiv M(\xi=\xi_{b}) = \frac{a_{0}^{3}}{(4\pi G^{3})^{1/2}}\Big(\xi^{2}\frac{d\psi}{d\xi}\Big)^{1/2}_{\xi_{b}}
\end{equation}
(Bonnor 1956, Ebert 1955). Here $G$, is the gravitational constant, $\xi=\frac{r}{R_{0}}$, is the dimensionless radius of the {\small BE} sphere where $R_{0}=\Big(\frac{a_{0}^{2}}{4\pi G\rho_{c}}\Big)^{1/2}$, is the physical scale-factor for length and, $a_{0}$, the isothermal sound-speed. The central density, $\rho_{c}$, for the {\small BE} sphere is defined as
\begin{equation}
\rho_{c} \equiv \frac{a_{0}^{6}}{4\pi G^{3}M^{2}}\Big(\xi^{2}\frac{d\psi}{d\xi}\Big)^{2}_{\xi_{b}},
\end{equation}
so that the radial distribution of density within the {\small BES} is
\begin{equation}
\rho(r) = \rho_{c}\mathrm{e}^{-\psi(\xi)},
\end{equation}
 $M\equiv M(\xi)$, and  $\psi$ is the well-known Lane-Emden function (Chandrasekhar 1939). \\ \\
Now, Eqn. (1) can be used to calculate the magnitude of gas temperature required to support a core modelled as a {\small BES} of radius, $\xi_{b}$, and having mass, $M_{core}$, against self-gravity as
\begin{equation}
T^{exp}_{gas}(\xi_{b}) = \frac{\bar{m}}{k_{B}}\Big[M_{core}(4\pi \rho_{c}G^{3})^{1/2}\Big(\xi\frac{d\psi}{d\xi}\Big)^{-1}_{\xi_{b}}\Big]^{2/3},
\end{equation}
$\bar{m}$, being the mean molecular mass of gas. Finally, the magnitude of the pressure, $P_{ext}$, confining the {\small BES} is simply
\begin{equation}
P_{ext}\ \equiv\ P(\xi=\xi_{b}).
\end{equation}

\subsection{Numerical algorithm}
Simulations discussed in this work were developed using the Smoothed particle hydrodynamics ({\small SPH}) code, {\small SEREN} (Hubber \emph{et al.} 2011). {\small SEREN}  is a well-tested, state-of-the-art {\small SPH} code that incorporates all the basic features of this algorithm. The calculation proceeds by stacking particles representing the model core on an octal tree that is used to identify the nearest neighbours of a particle(exactly 50), and to calculate the magnitude of net force experienced by a particle. The standard cubic-spline kernel is employed to smooth out contributions from individual particles and the smoothing length of each particle is adjusted such that it has exactly 50 neighbours. The prescription of the Riemann artificial viscosity with the viscosity parameter, $\alpha$=0.5, suggested by Monaghan (1997) was preferred in this set of simulations. \\
\subsection{Calculating dust and gas temperature} Temperature of individual gas particles in this work was calculated by explicitly solving the respective equations of thermal balance for the gas and dust. Dust in the cores is primarily heated by the interstellar radiation field ({\small ISRF}), characterised in the present work by a heating rate, $\Gamma_{{\small ISRF}}$. Other functions that characterise the thermal energy budget of a dust grain are its cooling function, $\Lambda_{dust}$, and gas-dust coupling function, $\Lambda_{g-d}$, that accounts for energy transfer during collisional interaction between a dust grain and a gas molecule. $\Lambda_{g-d}$, is effectively a cooling-function when dust is cooler than gas as it gains thermal energy  from warmer gas molecules. On the contrary, $\Lambda_{g-d}$, behaves like a heating-function when dust is warmer than gas. In the present work we use the standard heating/cooling rates suggested by Goldsmith (2001) for typical clouds in the Solar neighbourhood, and where applicable, those corrected elsewhere in literature. The dust heating-rate due to the {\small ISRF} 
\begin{equation}
\Gamma_{ISRF} = 3.9\times 10^{-24}\Big[\frac{n(H_{2})}{\mathrm{cm}^{-3}}\Big]\chi \ \mathrm{erg}\mathrm{cm}^{-3}\mathrm{s}^{-1},
\end{equation}
where, $\chi\sim 10^{-4}$, is the factor by which the {\small ISRF} is attenuated in a putative star-forming clump, while $n(H_{2})$ is the molecular hydrogen number density. We hold this magnitude of $\chi$ fixed over the entire duration of a simulation. This quantity is reasonably well constrained in literature; Valdivia et al. (2016) showed that $\chi\sim 10^{-4}$ for typical prestellar densities($\gtrsim 10^{4}$ cm$^{-3}$) which is also consistent with the magnitude reported for {\small IRDCs}\footnote{Infrared Dark clouds.} (Goldsmith 2001). \\ \\
Next,
\begin{equation}
\Lambda_{g-d} = 2\times 10^{-33}\Big[\frac{n(H_{2})}{\mathrm{cm}^{-3}}\Big]^{2}\Big(\frac{\triangle T}{K}\Big)\Big(\frac{T_{gas}}{10 K}\Big)^{0.5} \ \mathrm{erg}\mathrm{cm}^{-3}\mathrm{s}^{-1},
\end{equation}
for standard values of dust-to-gas ratio and dust grain-size (Burke \& Hollenbach 1983); $\triangle T = T_{dust} - T_{gas}$, $T_{dust}$ and $T_{gas}$ being respectively the temperature of dust, and gas. In one of the test cases, listed 8 in Table 1, we raised the dust grain-size by about a factor of eight to mimic the effect of the so-called fluffy dust-grains. Fluffy dust-grains enhance the contribution due to $\Lambda_{g-d}$ and assist gas-cooling when the dust is cold. This is usually true for cores in typical low-mass star-forming clouds. In fact, we then repeated calculations for all the other test cores(listed 9-12 in Table 1) to investigate if the puffed-up dust grains also lowered the gas-temperature in other starless cores before eventually leading them to collapse. Finally, the dust cooling function,
\begin{equation}
\Lambda_{dust} = 4.22\times 10^{-31}\Big(\frac{T_{dust}}{K}\Big)^{6}\Big[\frac{n(H_{2})}{\mathrm{cm}^{-3}}\Big] \ \mathrm{erg}\ \mathrm{cm}^{-3}\ \mathrm{s}^{-1}
\end{equation}
(Bate \& Keto 2015). Without an elaborate chemical network at our disposal cooling of molecular gas was implemented by adopting a parametrised cooling function, $\Lambda_{gas}$, for undepleted molecular abundances in clumps,
\begin{equation} 
\Lambda_{gas} = A\Big(\frac{T_{gas}}{10 K}\Big)^{B}
\end{equation}
Goldsmith(2001); this cooling function mimics very well the observed cooling-rate in clumps within the Solar neighbourhood. The coefficients $(A,B)$ here are the same as $(\alpha,\beta)$ used by Goldsmith (2001) and listed in his Table 1. We avoid $(\alpha,\beta)$ here for fear of confusing them with the {\small SPH} artificial viscosity coefficients.
We assume that gas in a test core is heated externally only due to cosmic-rays({\small CRs}), for it is reasonable to neglect photoelectric heating since cores are usually found in well shielded regions of a molecular cloud. The heating rate, $\Gamma_{CR}$, due to {\small CRs} is,
\begin{equation}
\Gamma_{CR} = 10^{-27}\Big[\frac{n(H_{2})}{\mathrm{cm}^{-3}}\Big] \ \mathrm{erg}\ \mathrm{cm}^{-3}\ \mathrm{s}^{-1}. \\ \\
\end{equation}
Finally, the equilibrium dust and gas temperature was calculated by solving simultaneously the respective equations of thermal balance,
\begin{eqnarray}
\Gamma_{{\small ISRF}} - \Lambda_{dust} + \Lambda_{g-d} = 0 \\
\Gamma_{CR} - \Lambda_{gas} - \Lambda_{g-d} = 0.
\end{eqnarray}
The equilibrium gas temperature, $T_{gas}^{eq}$, of a particle, $p$, corresponds to an equilibrium energy, $u_{p}^{eq}$, that is used to define the thermal timescale, $dt_{thermal}$, for an {\small SPH} gas particle as
\begin{equation}
dt_{thermal} = \vert u_{p} - u_{p}^{eq}\vert\Big(\frac{du}{dt}\Big)_{net}^{-1},
\end{equation}
where $\Big(\frac{du}{dt}\Big)_{net}$ is the net heating/cooling that the particle is subject to. It is the equivalent of $(n^{2}\Lambda - n\Gamma)$; $\Lambda$ and $\Gamma$ being  respectively the effective cooling, and heating rate for an {\small SPH} particle. The internal energy of a particle was then revised according to
\begin{equation}
u_{p}' = u_{p}^{eq} + (u_{p} - u_{p}^{eq})\exp\Big(\frac{-dt}{dt_{thermal}}\Big),
\end{equation}
where $dt$ is the real time-step for the particle $p$. In case a particle is heating/cooling in quasistatic manner, $dt_{thermal}\gtrsim dt$, so that the above equation reduces to
\begin{equation}
u_{p}' = u_{p} - dt\Big(\frac{du}{dt}\Big)_{net}.
\end{equation}
On the other hand if it cools/heats rapidly, $dt_{thermal}\ll dt$, so that it quickly attains its equilibrium energy and $u_{p}' = u_{p}^{eq}$. Temperature, $T_{ext}$ (see Table 1), of the {\small ICM} particles confining a model core was held constant throughout the length of a realisation. We must note that our extant interest is confined only to the low-mass star-forming clouds in the local Universe. The question about the impact of the ambient environment on the evolution of cores, though pertinent, is beyond the scope of this work. It must also be added that the choice of the respective heating functions $\Gamma_{CR}$, $\Gamma_{ISRF}$ and the molecular-line cooling function, $\Lambda_{gas}$, used in the calculations presented in this work is appropriate, for the external heating is significantly stronger in regions of high mass star-formation and at high redshifts (see e.g. Redman \emph{et al.} 2004). Also, the details of molecular cooling are likely to be different in these respective environs. 

\subsection{Initial conditions}
The choice of initial conditions is relatively simple and consists of a core in approximate (thermal)pressure equilibrium with its confining medium. All test cores were initially Jeans stable. As discussed in \S 2.1 we modelled each of our test core as an isothermal pressure-confined {\small BES}($\xi_{b}$=3), with the exception of the core {\small B68} that was also modelled as a critically stable {\small BES}($\xi_{b}$=6.5) and a uniform density sphere to test if the specific choice of the initial density distribution possibly bears upon the evolution of the core. The magnitude of temperature, $T_{gas}$, initially assigned to the gas in a model core  was obtained from literature for the respective core. The initial radius of the core, $R_{core}$, listed in column 4 of Table 1 was calculated using $T_{gas}$ in Eqn. (4) above that ensures the initial set-up is in pressure equilibrium. The analogous expression for a core assembled as a sphere of uniform density is 
\begin{displaymath} 
T_{gas} = \frac{\bar{m}G M_{core}}{3k_{B}R_{core}},
\end{displaymath}
where symbols have their usual meaning. As noted earlier, particles representing the externally confining {\small ICM} were always maintained at a fixed temperature, $T_{ext}$, which was calculated using Eqn. (5) for the respective BESs while for the Uniform density sphere, it was calculated by assuming the same centre-to-boundary density contrast as that for a pressure-confined BES($\xi_{b}$ = 3). Listed in Column 7 of Table 1 is the ratio of the energy due to self-gravity against that due to the externally confining medium, $Q_{init}$; a magnitude smaller than unity implies a pressure-confined configuration. Each of the three polytropes were assembled by randomly positioning  particles within them. The particle distribution in each polytrope was relaxed by allowing it to evolve isothermally for a few sound-crossing times. The resulting density distribution for the respective {\small BES}s has been shown on the two panels of Fig. 3. The settled polytropes were then stretched to the desired core radius, $R_{core}$. The model core in this work was assumed to be composed of the usual cosmic mixture.\\ \\
\textbf{Resolution} The number of particles,$N_{gas}$,  to be assembled in a polytrope was calculated using the equation
\begin{equation}
N_{gas} = \Big(\frac{3}{32\pi}\Big)\cdot\Big(\frac{N_{neibs}}{h_{avg}^{3}}\Big)R_{core}^{3},
\end{equation}
$N_{neibs}$ = 50, being the number of neighbours of each particle. The average smoothing length, $h_{avg}$, that determines the spatial extent of the smallest resolvable region in a realisation was calculated such that it satisfied the Truelove criterion (Truelove 1997), of resolving the Jeans instability. If $\lambda_{J}$ is the Jeans length at the minimum gas temperature to be attained ($\sim$7 K) in these realisations for the typical protostellar density, $\sim 10^{7}$ cm$^{-3}$, then the Truelove criterion demands, $h_{avg}\lesssim \frac{\lambda_{J}}{4}\sim 281$ AU, which ensures there is no artificial fragmentation at the desired density. Furthermore, this choice of spatial resolution transforms into a much higher resolution for the rest of the core since $\frac{R_{core}}{h_{avg}}\gg$ 30 for each realisation. Consequently, these simulations not only satisfy the {\small SPH}  equivalent of the Truelove criterion (Hubber \emph{et al.} 2006), but also the more stringent criterion suggested by Federrath \emph{et al.} (2014) that ensures convergence in the calculation of energy and momentum within a test core. Simulations reported here were developed with $\sim$0.7 M particles out of which $\sim$0.3 M represented the {\small ICM}. Each realisation required about 1300 {\small CPU} hours with 8 threads. \\ \\
Apart from these, $N_{icm}$ number of particles representing the externally confining {\small ICM} and always maintained at temperature, $T_{ext}$,  were assembled in a jacket of thickness of 2$h_{avg}$ around the core. The number of {\small ICM} particles was calculated such that the number density of particles on either side of the gas-{\small ICM} interface was identical.  The core+{\small ICM} assembly was then enveloped by a layer of dead particles, the so-called boundary particles, to prevent gas/{\small ICM} particles from escaping. The initially static {\small ICM} particles exert only the hydrodynamic force on gas particles while the boundary particles do not contribute to the net-force and always remain static in the computational domain. \\ \\
\begin{figure*}
\vspace{1pc}
\includegraphics[angle=270,width=\textwidth]{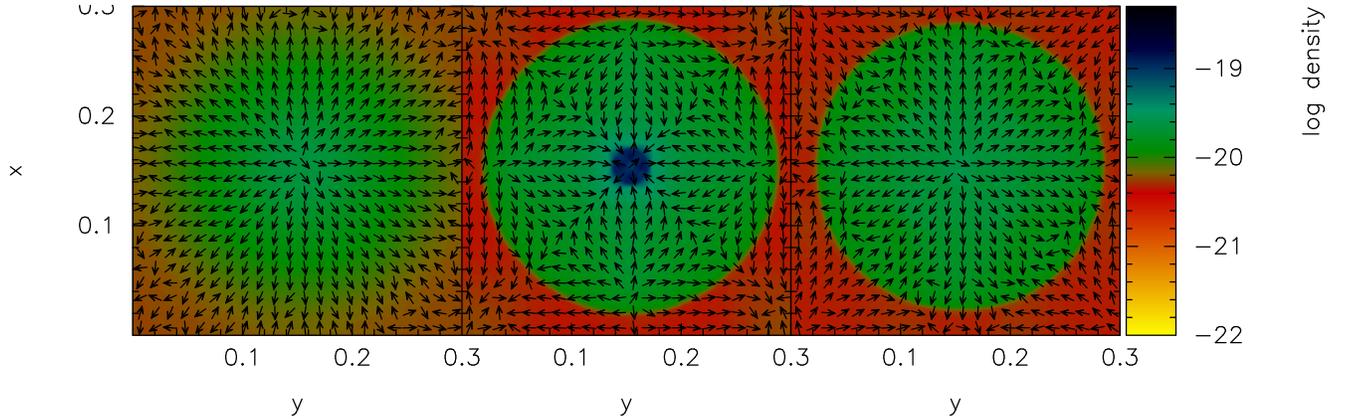}
\caption{Rendered density plots for the model B68 core, listed 1 in Table 1, show the assembly of a centrally condensed core due to an inwardly directed compressional wave. Epochs corresponding to the images on respective panels of this figure are $t$=0.001 Myr, 0.32 Myr and 1 Myr; arrows overlaid on these plots represent the direction of the underlying gas-flow.}
\end{figure*}

\textbf{Representing the protostar in case the model core became singular} \\
The protostar in a realisation was represented by a sink particle, the density threshold, $n_{sink}$, for which was set at $n_{sink}\sim\ 10^{7}$ cm$^{-3}$. In other words, an {\small SPH} particle with density exceeding the threshold, $n_{sink}$, was replaced with a sink particle if in addition it also had a negative divergence of velocity and acceleration (e.g. Bate \emph{et al.} 1995; Bate \& Burkert 1997; Federrath \emph{et al.} 2010). The radius of a sink particle was set at 2.5 times the average {\small SPH} smoothing length of the candidate sink particle and turns out to be on the order of $\sim 800$ AU. Gas particles that enter the sink radius were accreted by it.

\section{Results}
\subsection{Starless cores investigated in this work} 
In order to justify the choice of initial conditions listed in Table 1, we begin by briefly examining the physical properties of these cores documented in contemporary literature. This endeavour will also serve us well while testing the validity of results deduced from our realisations. \\ \\
\textbf{{\small B68}} One of the most well-studied cores and is reported to have a density profile that matches the density distribution of a unstable {\small BES} of radius, $\xi_{b}$ = 6.9 and radius, $r_{c}\sim$ 0.05 pc (e.g. Alves \emph{et al.} 2001), though there is a slight variation in the estimate of its radius (between $\sim$0.035 pc - 0.05 pc), depending on the assumed distance to the core (e.g. Nielbock \emph{et al.} 2012; Schnee \emph{et al.} 2013). \\ \\
\textbf{{\small L694-2}} First classified as starless by Lee \& Myers (1999) and Lee \emph{et al.}(1999), using data from the {\small SDSS} survey. Estimates of its physical parameters vary and a mass between 0.5 M$_{\odot}$ (Larshakov \emph{et al.} 2013, by studying NH$_{3}$ emission) and 3.1 M$_{\odot}$ (Harvey \emph{et al.} 2003) has been suggested, subject to the tracer used to study it. Similarly, the core has been suggested to have a radius between 0.05 pc (Williams \emph{et al.} 2006) and 0.15 pc (Harvey \emph{et al.} 2003). In a more recent contribution, Chitsazzadeh   (2014) suggested a core mass of $\sim$2.5 M$_{\odot}$ and a radius $\sim$0.1 pc and furthermore, their radiative transfer modelling of {\small L694-2} showed that its cold interior was cocooned by a warmer envelope. They estimated the gas temperature to be $\sim$19 K in the envelope of this core and $\sim$ 7 K towards its centre. \\ \\
\textbf{{\small L1517B}} A starless core that is probably experiencing a breathing mode, i.e., exhibiting signs of in-fall close to the centre, but has an expanding envelope (Keto \& Field 2005; Fu \emph{et al.} 2011). As with other cores, estimates of its physical properties are subject to the molecular tracer. For instance, the mass and radius for this core varies between $\sim$0.05 M$_{\odot}$ and $\sim$0.06 pc(Benson \& Myers 1989) to $\sim$1.7 M$_{\odot}$(Kirk \emph{et al.} 2005) and  $\sim$3.89 M$_{\odot}$(Fu \emph{et al.} 2011). The gas temperature for this core has been reported to be $\sim$9 K(Tafalla \emph{et al.} 2002; Keto \& Field 2005). \\ \\
\textbf{{\small L1689-SMM16}} A unusually massive starless core  and supposedly has mass well in excess of its (thermal)Jeans mass, $M_{Jeans}$, and therefore has been described as \emph{Super-Jeans} (Sadavoy \emph{et al.} 2010a,b). However, signs of outwardly directed gas-flow have been detected in its envelope, but is believed to be on the verge of collapse given the relatively large, $\frac{M_{core}}{M_{Jeans}}$, ratio (Chitsazzadeh \emph{et al.} 2014).\\ \\
\textbf{{\small L1521F}}  Probably associated with a weak, poorly collimated outflow and is believed to have formed its first hydrostatic core. Such objects have a relatively small luminosity, on the order of $\lesssim$0.1 L$_\odot$, and are known as Very low luminosity objects({\small VeLLOs}) in contemporary literature (e.g. Di Francesco \emph{et al.} 2007). Some of the earliest detections of such objects were reported by Young \emph{et al.}(2004) (e.g. {\small L1014-IRS}) and Kauffmann \emph{et al.}(2005)(e.g. L1148-IRS). The central condensation in {\small L1521F} is sub-stellar and has a bolometric luminosity of $\sim$0.05 L$_{\odot}$ (e.g.  Bourke \emph{et al.} 2006; Terebey \emph{et al.} 2009), and estimates of mass for this core vary between 3 M$_{\odot}$ (Onishi \emph{et al.}1999) and 5.5 M$_{\odot}$ (Crapsi \emph{et al.} 2004). In the present work we adopt the more recently deduced physical properties for this core (Chitsazzadeh 2014).  Radiative transfer models developed by this author for the {\small L1521F} constrained the temperature of gas in the envelope to between 18 K and 21 K.

\subsection{Evolution of gas density}
The polytrope in each realisation became centrally concentrated due to an inwardly propagating compressional wave triggered by the gradual cooling of gas in it. In the set of realisations discussed here the pressure exerted by the confining {\small ICM} was initially purely thermal in nature since the particles resembling the {\small ICM} were initially static. The initial magnitudes of $P_{ext}$ are therefore likely to be somewhat smaller, perhaps by a factor of a few, in comparison with the magnitude of pressure exerted by the {\small ICM} in typical star-forming clouds. Observationally, however, it is known that a non-thermal component also makes a significant contribution to the pressure, $P_{ext}$ (e.g Bailey \emph{et al.} 2014), in these clouds. The somewhat lower initial magnitude of $P_{ext}$ is unlikely to affect our calculations here since the particles representing the {\small ICM} in these realisations are mobile during a calculation and therefore additionally contribute to $P_{ext}$ over the length of a calculation. \\ \\
\begin{figure}
\vspace{1pc}
\includegraphics[angle=270,width=0.5\textwidth]{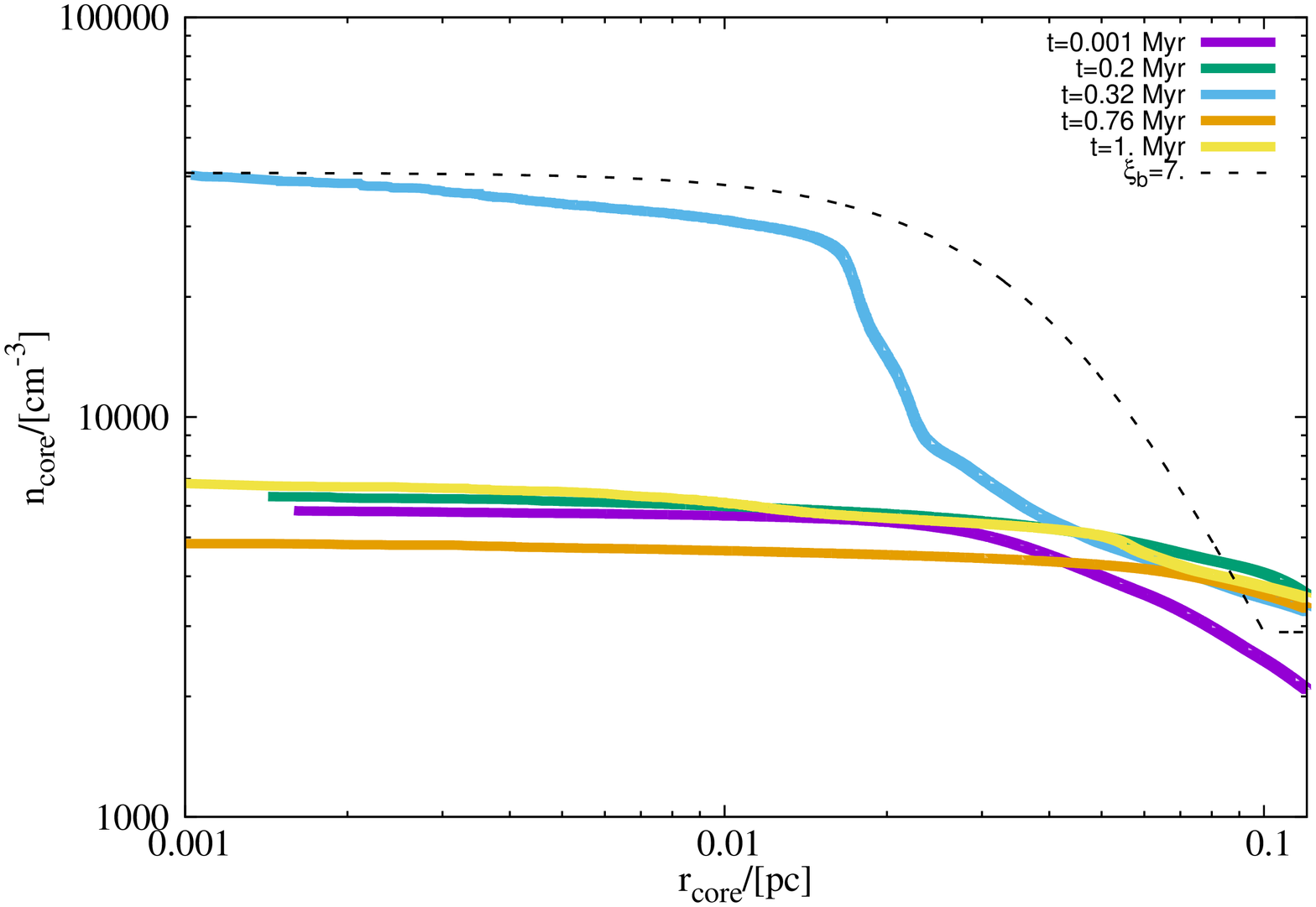}
\includegraphics[angle=270,width=0.5\textwidth]{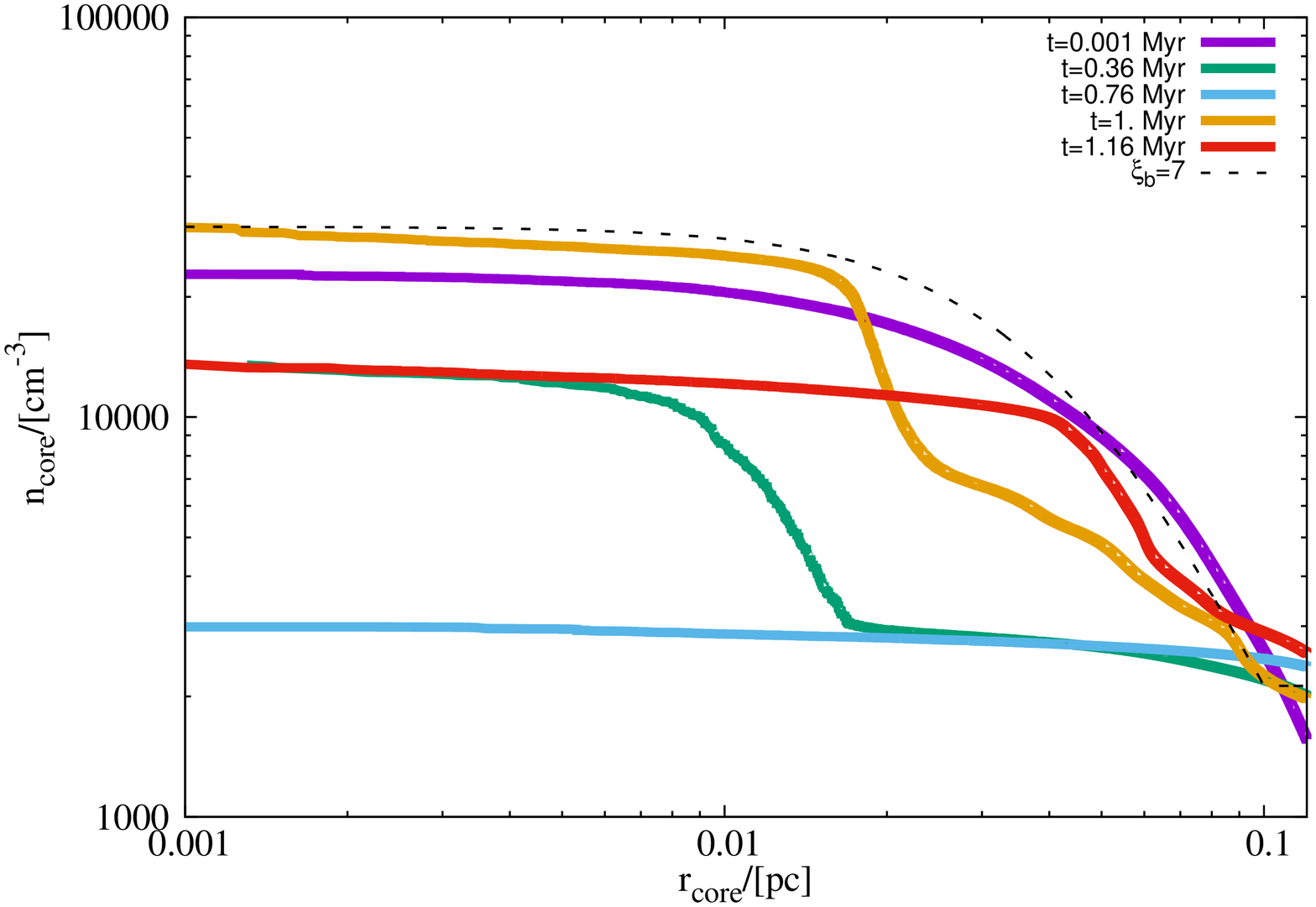}
\includegraphics[angle=270,width=0.5\textwidth]{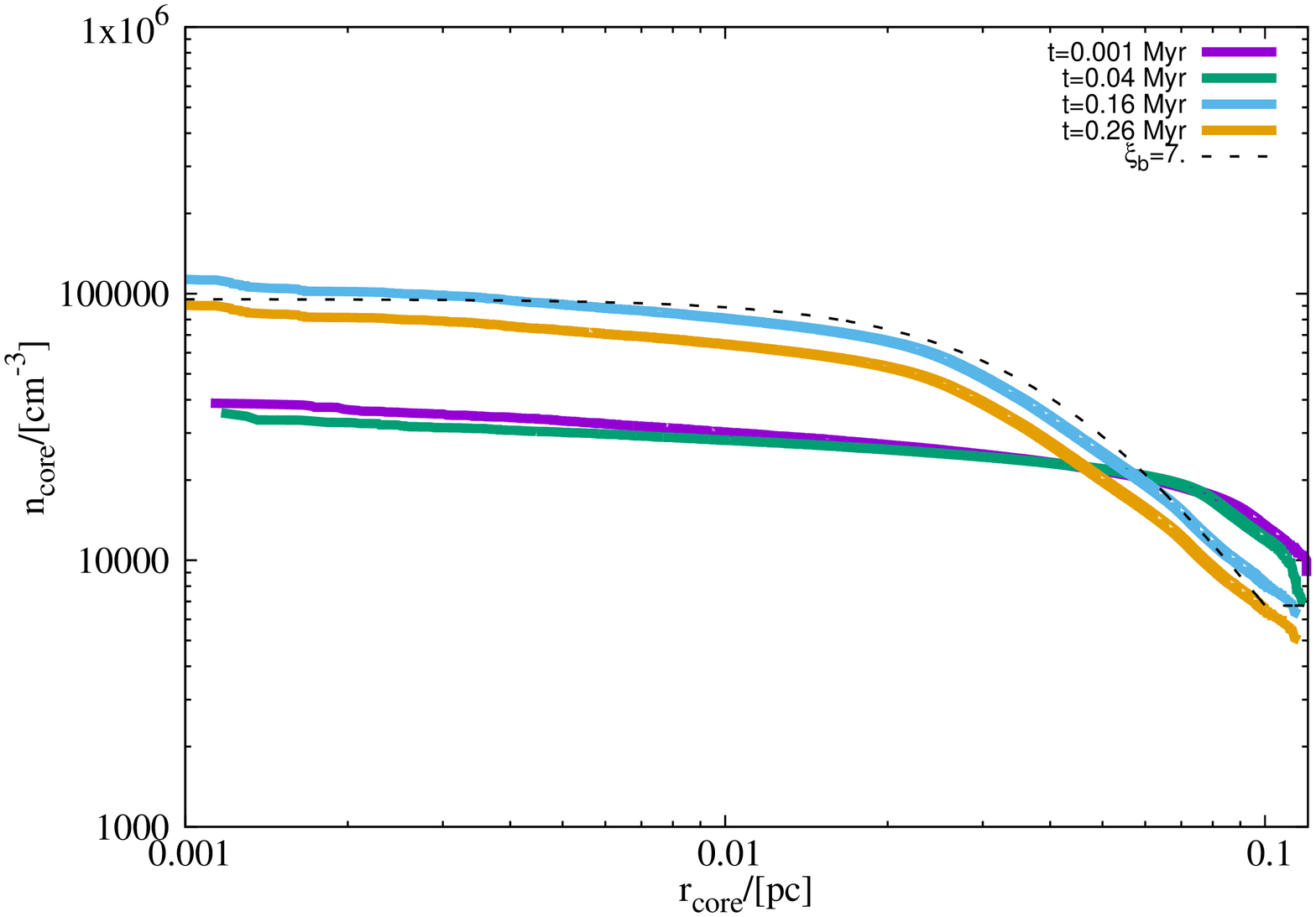}
\caption{Shown on the upper, central and the lower-panel are plots of the radial distribution of gas density at different epochs of the core B68 modelled as respectively, the stable Bonnor-Ebert sphere($\xi_{b}$=3), the critically stable Bonnor-Ebert sphere($\xi_{b}$=6.45) and a uniform density sphere. In each of these three realisations, the model core acquires a centrally condensed form when its density profile mimics that of a unstable Bonnor-Ebert sphere plotted with a dashed-black line, though the epoch at which this happens depends on the initial model of the core (\emph{see text}).}
\end{figure}
\begin{figure}
\vspace{1pc}
\includegraphics[angle=270,width=0.5\textwidth]{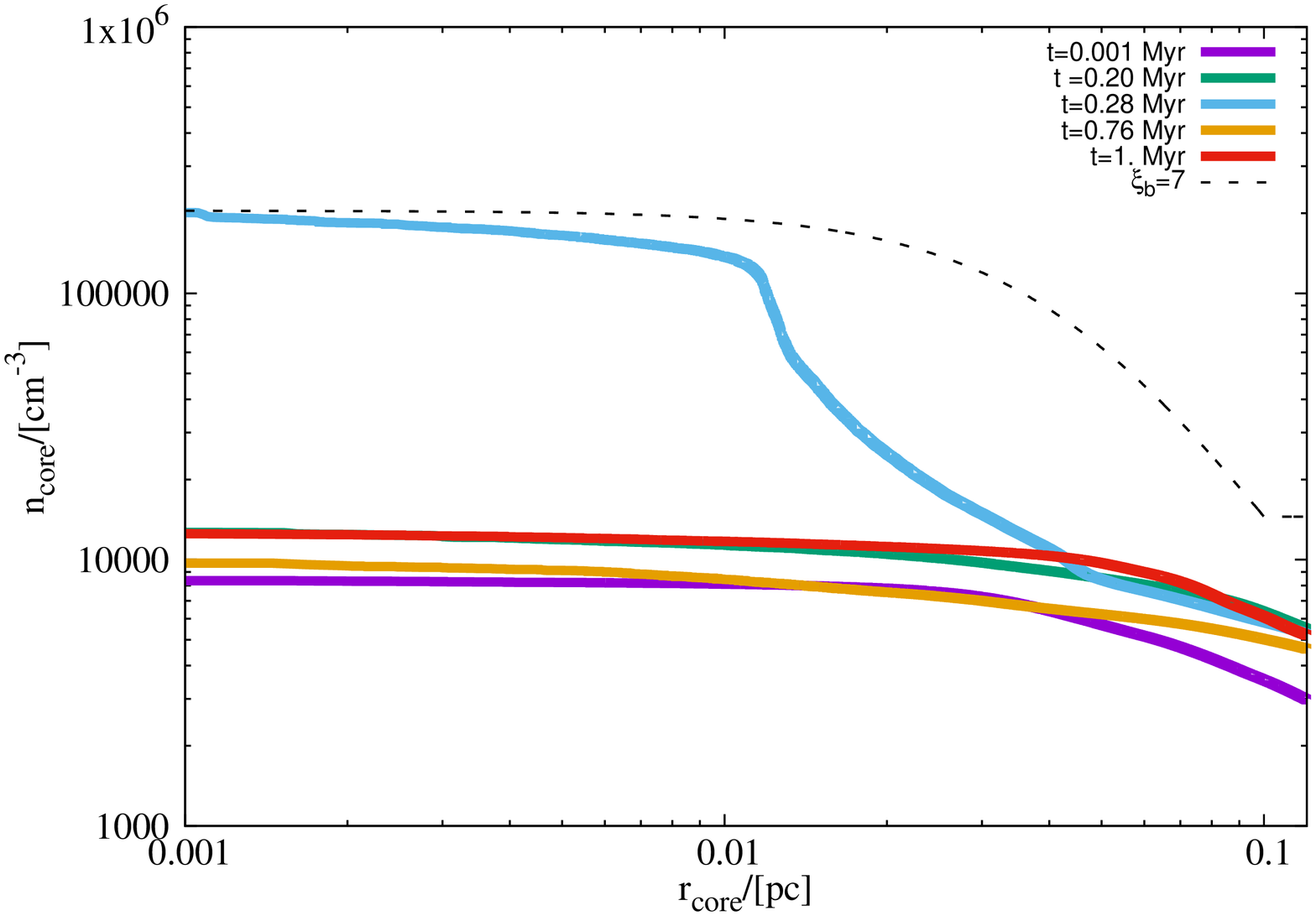}
\includegraphics[angle=270,width=0.5\textwidth]{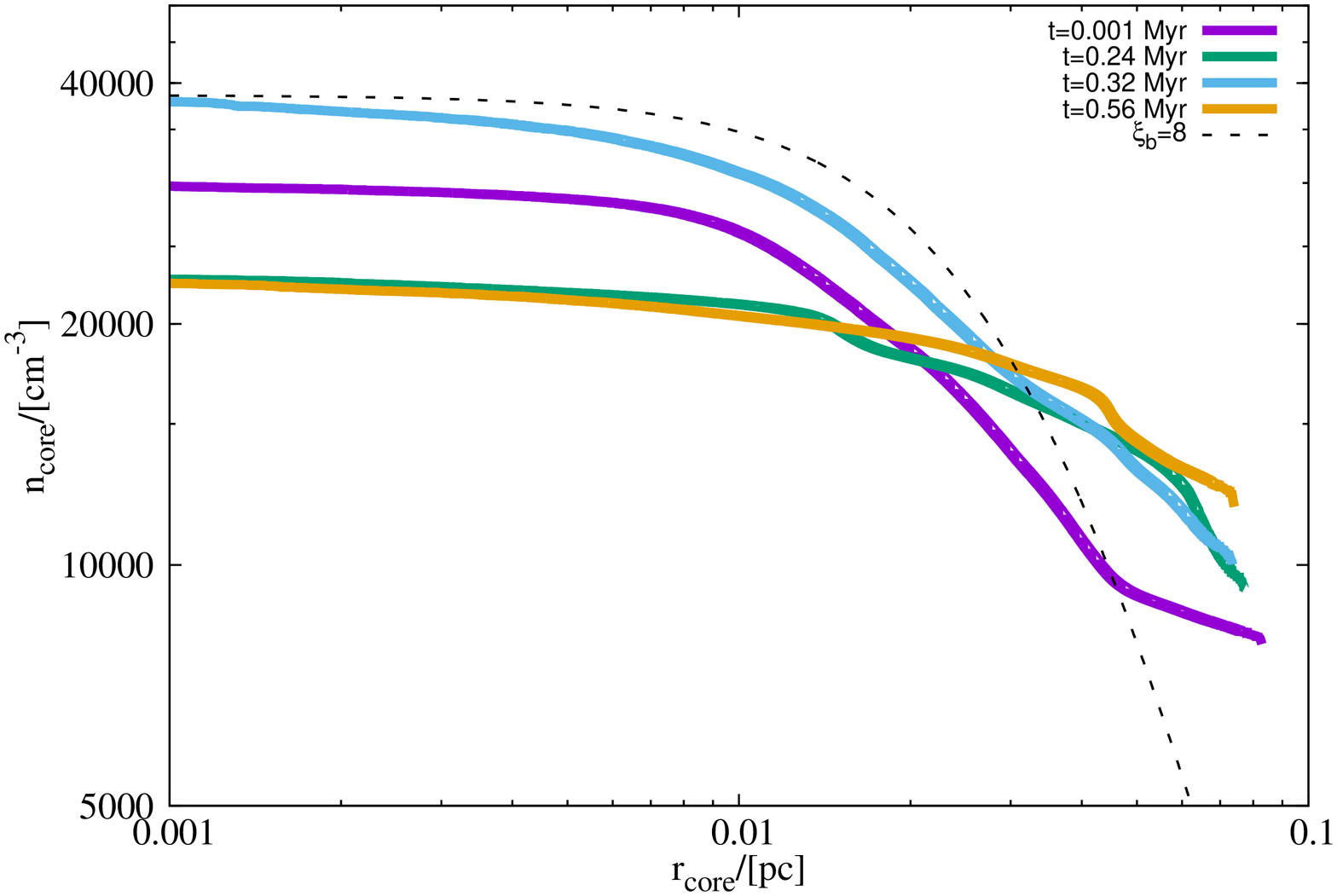}
\includegraphics[angle=270,width=0.5\textwidth]{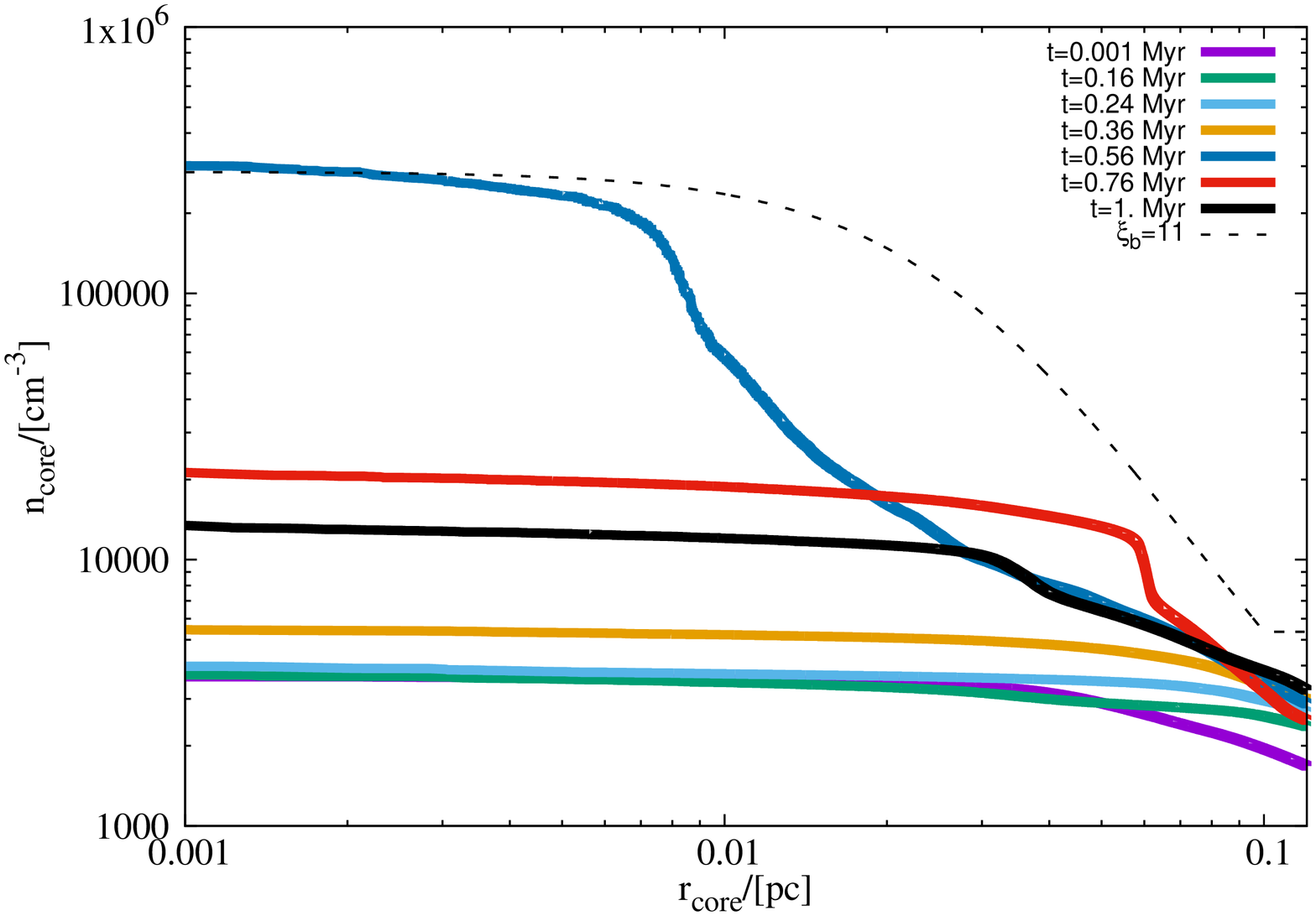}
\caption{Shown on the upper, central and the lower panel are the plots showing the radial distribution of gas density at different epochs for respectively the cores L694-2, L1517B, and L16189-SMM16. Note that each one them acquired a centrally condensed form whence their respective density profile mimicked that of a unstable BES. The epoch when each core acquires its peak density and the radius, $\xi_{b}$, of the BES that fits its density distribution is of course different.}
\end{figure}
\begin{figure}
\vspace{1pc}
\includegraphics[angle=270,width=0.5\textwidth]{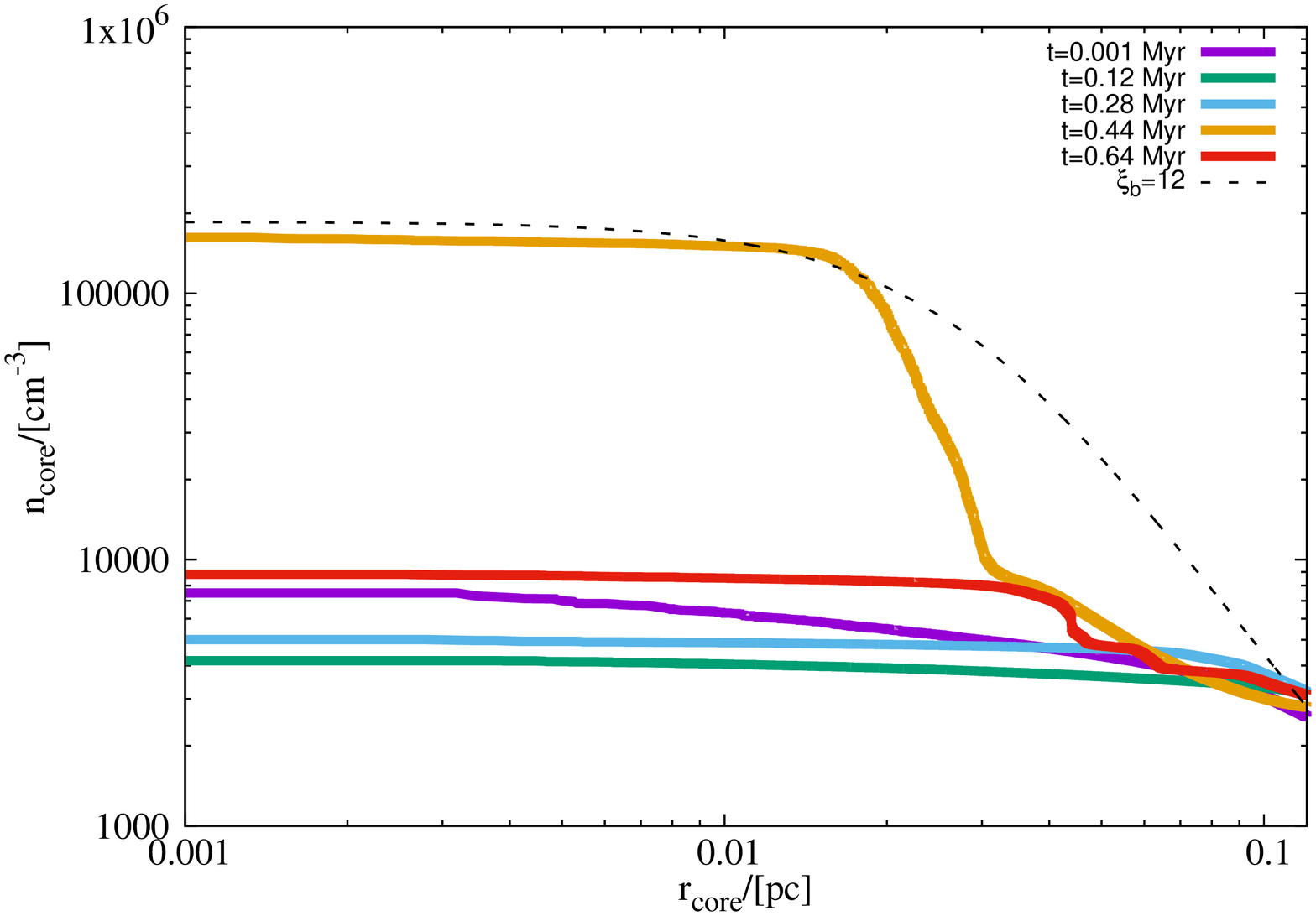}
\includegraphics[angle=270,width=0.5\textwidth]{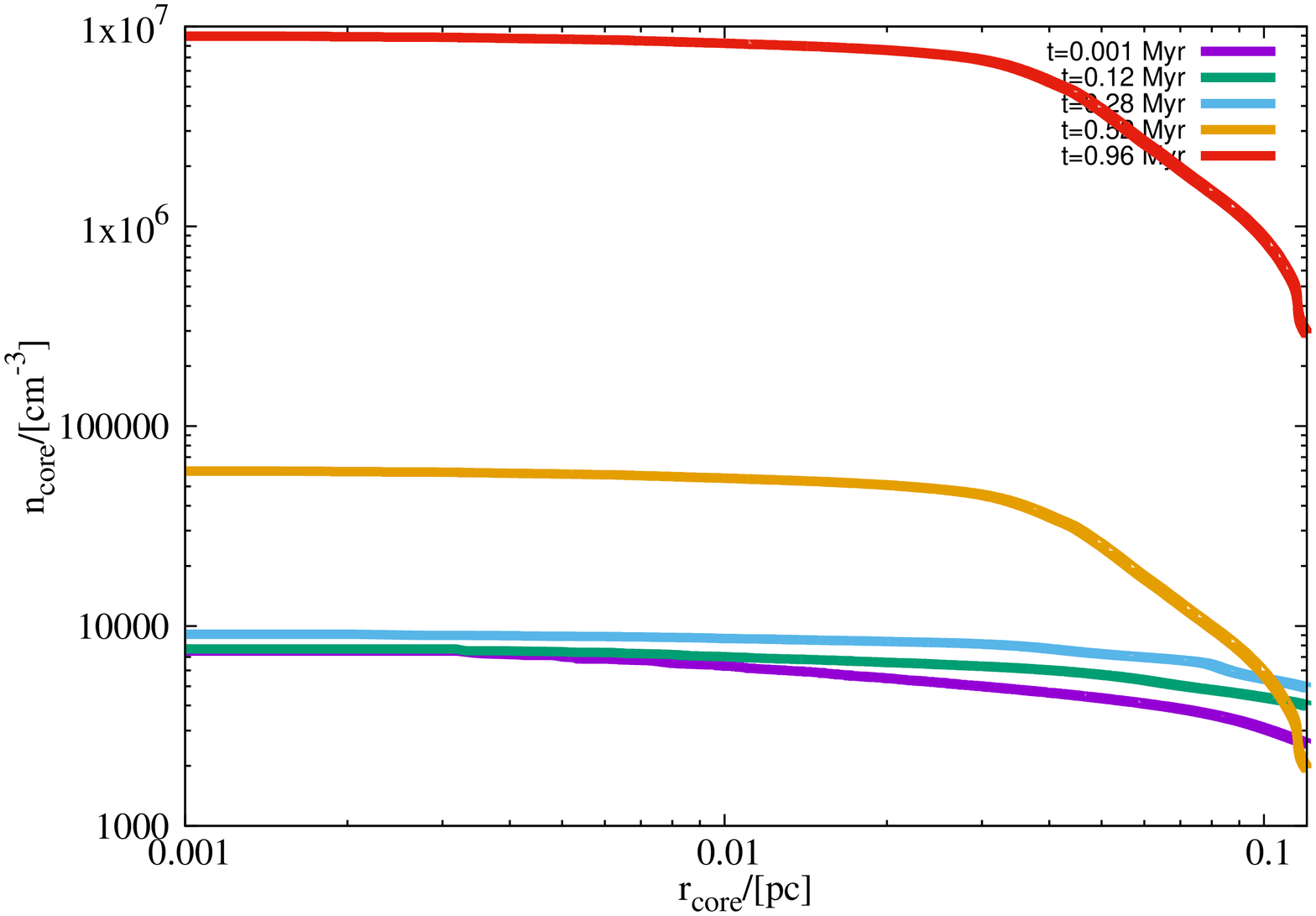}
\includegraphics[angle=270,width=0.5\textwidth]{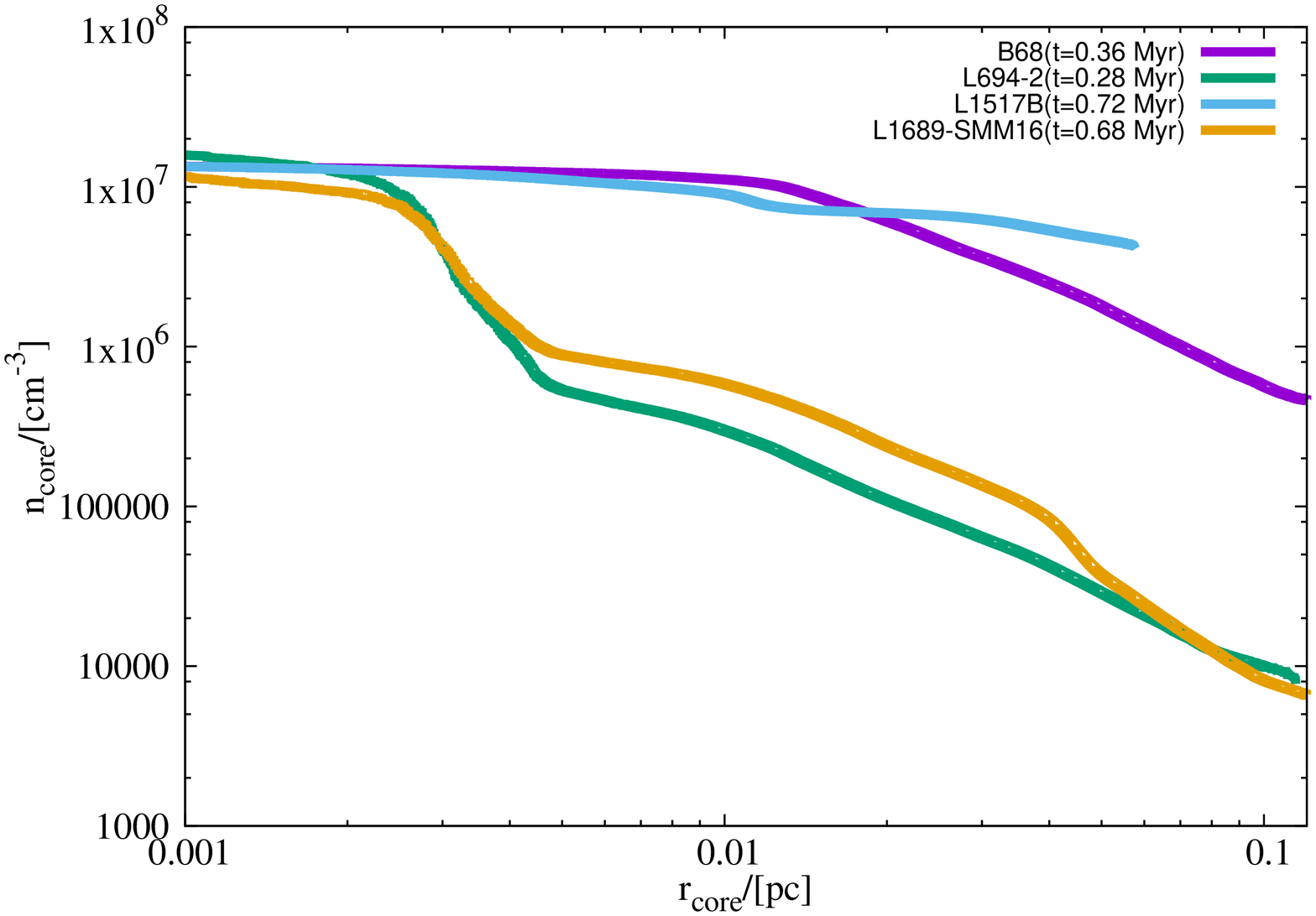}
\caption{Shown on the upper-panel of this plot is the radial distribution of gas density at different epochs of the core  L1521F modelled as a stable Bonnor-Ebert sphere($\xi_{b}$=3), and listed 7 in Table 1. The density distribution for the core that did in fact collapse to become a VeLLO, and listed 8 in Table 1, is shown on the central panel. Shown on the lower-panel is the radial density distribution for other cores listed 9-12 in Table 1 at the time respective calculations were terminated; \emph{see text for details}.}
\end{figure}
\textbf{Starless cores : \small{B68, L694-2, L1517B and L1689-SMM16}}
The picture in Fig. 4 is a representative plot showing the rendered density image of the cross-section of the mid-plane of the {\small B68} modelled as a pressure-confined {\small BES} ($\xi_{b}=3$), and listed 1 in Table 1. Having acquired its peak density, the core rebounded. This is evident from the picture on the central-panel of Fig. 4 that shows the epoch at which this core acquired its peak density and evidently, the inwardly directed compressional wave appears to have given way to expansion. Shown on the top-panel of Fig. 5 is the radial distribution of gas density within this core at different epochs. The model core is centrally most concentrated at $t\sim$ 0.32 Myr when in fact, its density profile is similar, or rather steeper than that of a unstable {\small BES} having radius, $\xi_{b}$ = 7. Also, at this epoch the radial density falls off by more than an order of magnitude over a radius, $r_{c}\sim$ 0.05 pc. We note that the core appears to be demonstrating breathing modes as it acquired a second density peak at $t\sim 1$ Myr when calculations for this realisation were terminated. The core {\small B68} was was also modelled as a critically stable {\small BES}($\xi_{b}\sim$ 6.5), and a sphere having uniform density. However, contrary to the previous realisation, the test core modelled as a critically stable {\small BES} initially showed expansionary tendency as is reflected by the initial fall in its central density. Then as the core continued to cool, and accompanied with a compressional wave, it became centrally peaked ($t\sim$ 1 Myr), with a density profile similar to that of a {\small BES} having radius $\xi_{b}$=7. Thereafter this test core rebounded as is evident from the plots shown on the central panel of Fig. 5. The uniform density sphere on the other hand, contracted rapidly and acquired a density peak at $t\sim 0.16$ Myr; see lower panel of Fig. 5. Qualitatively, the results from these latter two realisations converge with those deduced for the core modelled as a pressure-confined {\small BES}($\xi_{b}$=3). \\ \\
 As with the core {\small B68}, the other remaining cores viz., {\small L694-2, L1517B} and {\small L1689-SMM16} also became centrally concentrated as an inwardly propagating compressional wave swept them up. Shown in Fig. 6 are plots of the radial density distribution of gas in these latter cores at different epochs of their evolution. Observe that having acquired a centrally concentrated density distribution when the gas density in each core core mimicked the density profile of a unstable {\small BES}, the cores rebounded. The radius of each core at this epoch, i.e. the distance from the centre over which density falls over an order of magnitude, is consistent with various reported observations listed above. There is also evidence for radial oscillations in each of these realisations. \\ \\
\textbf{The core \small{L1521F}} 
The realisation listed 7 in Table 1 for this core was developed by adopting the standard dust grain-size in the gas-dust coupling function, $\Lambda_{g-d}$, given by Eqn. (7). The calculation was then repeated with dust grain-size an order of magnitude higher (listed 8 in Table 1). This effectively raised the contribution to gas cooling due to the collisional coupling between gas and dust.  The core in the former realisation did not collapse and like the starless cores discussed  previously, simply acquired a centrally peaked distribution before rebounding. In the latter case, however, the core did in fact become singular and formed a sub-stellar object represented by a sink particle. The temporal variation of the radial distribution of gas density in either realisation of this core is shown on respectively the upper and central panel of Fig. 7. Evidently, the higher dust grain-size proved efficient towards lowering the temperature of gas within the core which in turn assisted in-fall. Finally, shown on the lower-panel of this figure is the radial density profile for other test cores (realisations 9-12 in Table 1), but now with gas temperature calculated with puffed-up dust grains. The resulting density distribution for these respective cores is similar to that of the {\small VeLLO L1521F}. It is therefore clear that irrespective of the physical parameters of an individual core, a larger size of dust-grains leading to a higher contribution due to gas-dust coupling assists core-collapse. We will \textbf{expand} upon this further in \S 3.4 below where the temperature profile deduced for these cores will be presented.  \\ \\
Let us now examine the accretion history and the mass of the {\small VeLLO L1521F}. We remind, the central object in this test core was represented by a sink particle with a density threshold, $\sim 10^{7}$ cm$^{-3}$. This density threshold  is at least an order of magnitude higher than  the central density ($\sim 10^{6}$ cm$^{-3}$) reported by Onishi \emph{et al.}(1999)  and Crapsi \emph{et al.}(2004) for this core. Shown on the left-hand panel of Fig. 8 is the rate, $\dot{M}_{acc}\equiv\frac{dM_{*}}{dt}$,  at which the sink particle in this core accretes the in-falling gas. After an initially sluggish rate of accretion, it steadily gains momentum reaching a peak rate of $\sim 10^{-6}$ M$_{\odot}$ yr$^{-1}$, before eventually petering off over a period of $\sim 2\times 10^{5}$ years. This observed evolution of $\dot{M}_{acc}$ is far from uniform and in fact, is qualitatively consistent with the one derived by Whitworth \& Ward-Thompson (2001) for a core modelled with Plummer-like density profile.
Importantly, the magnitude of $\dot{M}_{acc}$ observed here is consistent with the observationally deduced magnitude by Takahashi \emph{et al.} (2013). Using this observed rate of accretion we can calculate the approximate mass of the central accreting object, $M_{*}$, with the expression for Bondi-Hoyle accretion
\begin{equation}
\dot{M}_{acc}=\frac{\bar{\rho}G^{2}M_{*}^{2}}{v_{r}^{3}},
\end{equation}
(Bondi 1952), where $\bar{\rho}\sim 10^{-17}$ g cm$^{-3}$, is the average density of in-falling gas, $v_{r}\sim$0.1 km/s and, $G$, the gravitational constant; other symbols have their usual meanings. The accretion-rate defined by Eqn. (17) gives a upper-limit to the magnitude of $\dot{M}_{acc}$; by contrast, the one in the Shu model of the collapse of a singular isothermal sphere, is constant (Anathpindika 2011). For an average accretion-rate, $\dot{M}_{acc}\sim\ 2\times 10^{-7}$ M$_{\odot}$ yr$^{-1}$, Eqn. (17) yields, $M_{*}\sim$0.04 M$_{\odot}$, a sub-stellar object, which is roughly consistent with the mass accreted by the sink-particle at the centre of this core as is visible from the plot shown on the right-hand panel of Fig. 8. Now, although the sink particle is likely to continue acquiring mass via accretion, the observed relatively low rate of accretion makes it likely that the object in this case will remain sub-stellar which is consistent with the fact that it is a {\small VeLLO}.
\begin{figure*}
\vspace{2pc}
\mbox{\includegraphics[angle=270,width=0.5\textwidth]{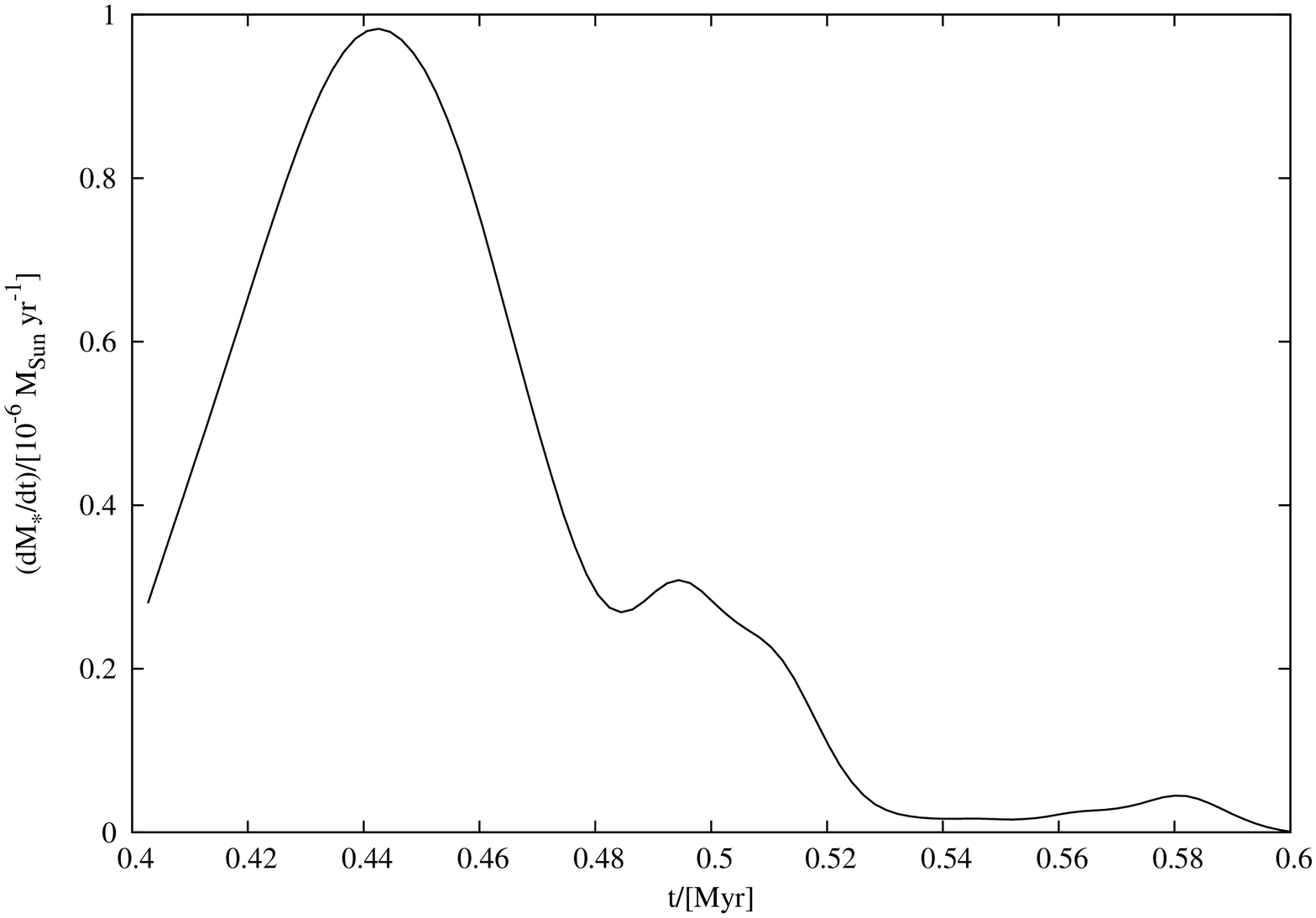}
\includegraphics[angle=270,width=0.5\textwidth]{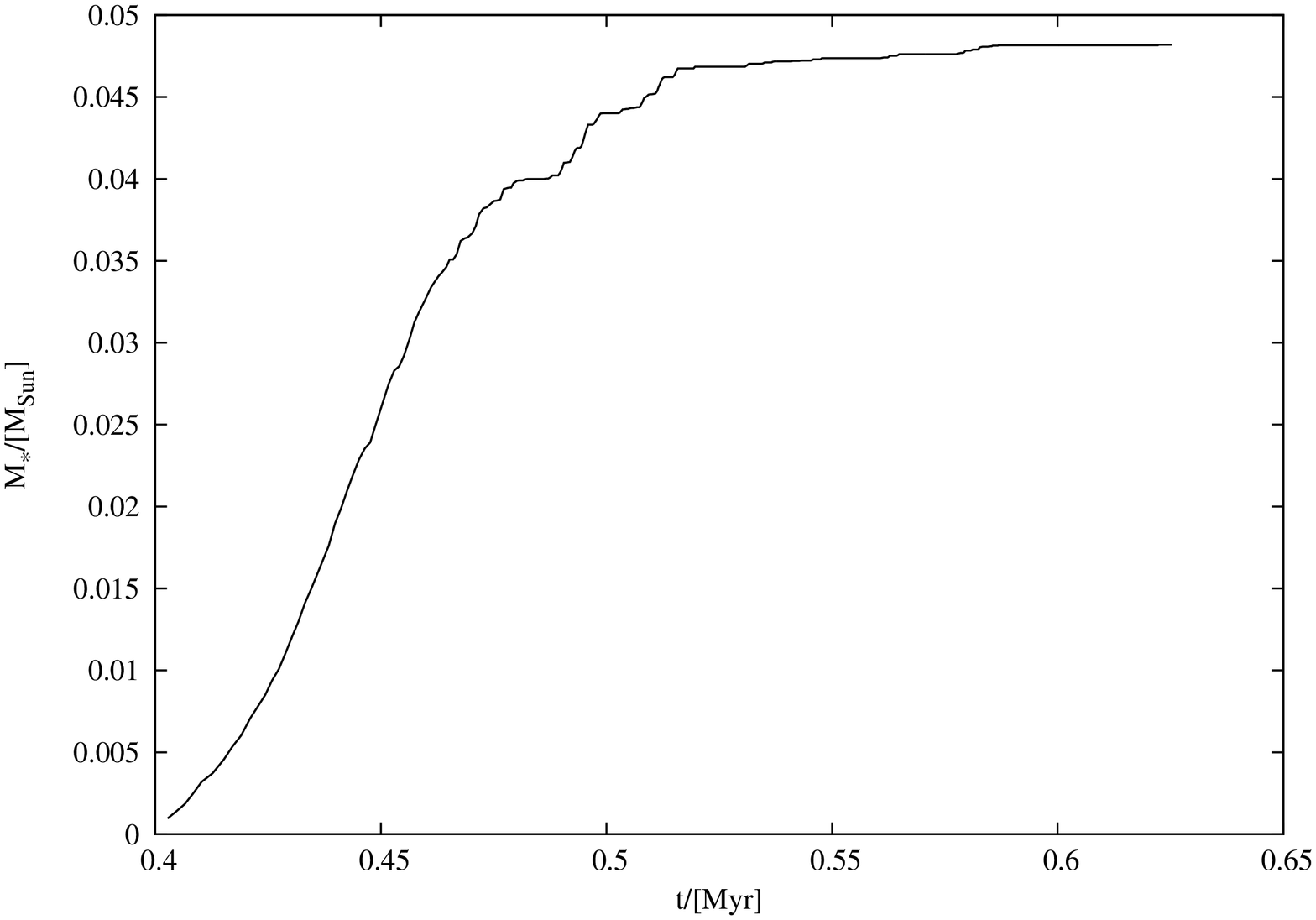}}
\caption{The plot on the left-hand panel shows the rate of mass-accretion of the sink particle that forms at the centre of the core whose density profile was shown on the central-panel of Fig. 7. The accretion-rate peaks at $\sim 10^{-6}$ M$_{\odot}$ yr$^{-1}$ before petering off. Shown on the right-hand panel of this figure is the total mass accreted by this sink particle.}
\end{figure*}


\subsection{Velocity field within contracting cores}
In the interest of brevity we will restrict the present discussion
only to the two contrasting cases viz., the starless core {\small B68} and the {\small VeLLO}, {\small L1521F}; the behaviour of the remaining starless cores was similar to that of the  core {\small B68}. Shown on the upper-panel of Fig. 9 is the radial component of gas velocity in this core. The negative magnitude of velocity on this plot as usual denotes inwardly directed gas-flow. Evidently, the onset of a strong compressional wave pushed gas inward that enhanced the density near the centre of the model core. Although this compressional wave can be seen to be the strongest just before the core achieved its central density peaks ($t\sim$0.20 Myr and $t\sim$ 0.76 Myr), the gas-flow within the core remained subsonic at all times. The magnitude of inward/outward gas-velocity is between $\sim$0.02 km/s - 0.15 km/s which is consistent with that inferred by for instance, Maret \emph{et al.} (2007) through observations of a number of emission lines towards {\small B68} and that suggested by radiative-transfer modelling of the core (e.g. Keto \emph{et al.} 2006). At early times, $t\sim 0.2$ Myr, gas within the interiors of the core appears to be outwardly directed before being swept-up by the compressional wave. \\ \\
\begin{figure}
\vspace{1pc}
\includegraphics[angle=270,width=0.5\textwidth]{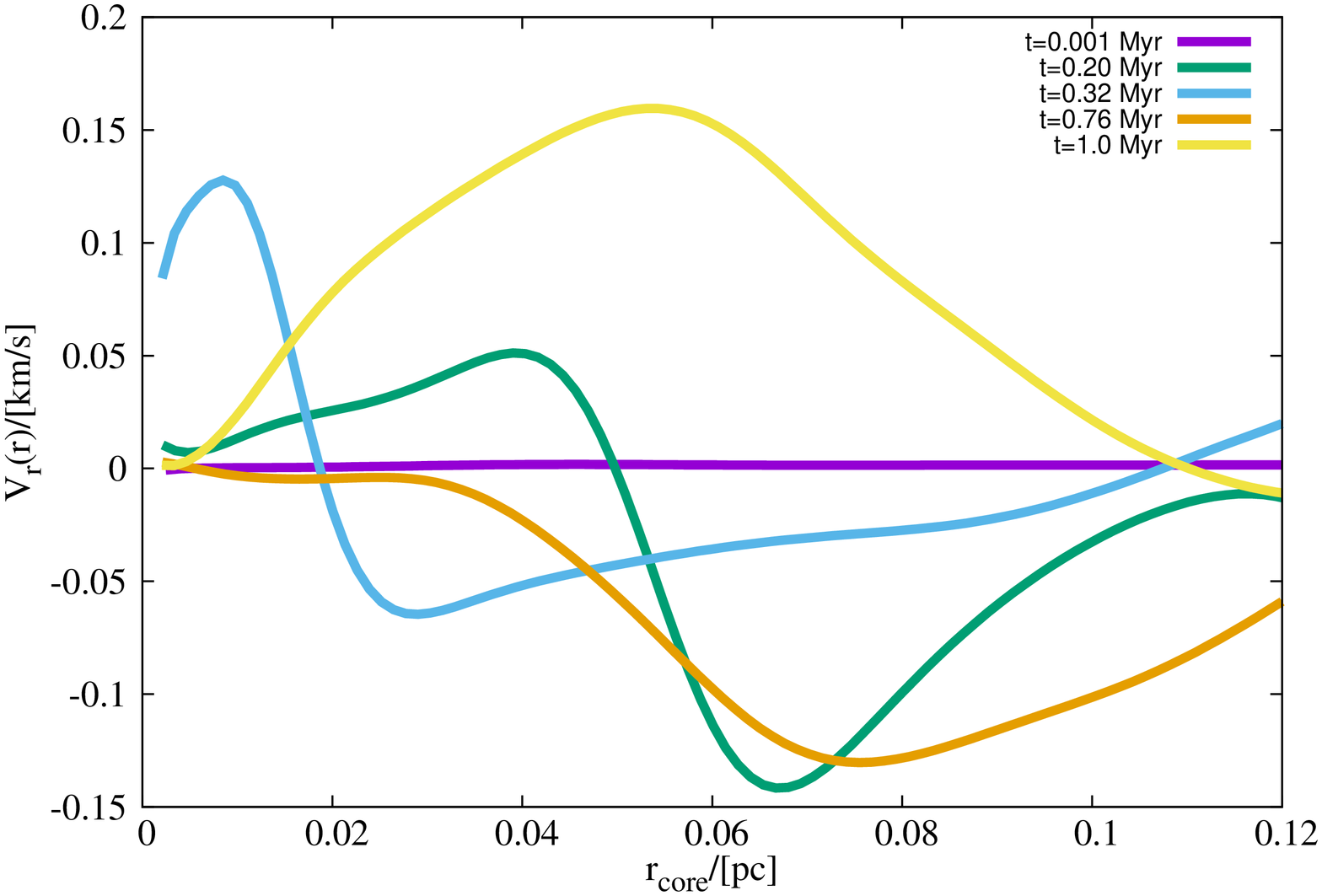}
\includegraphics[angle=270,width=0.5\textwidth]{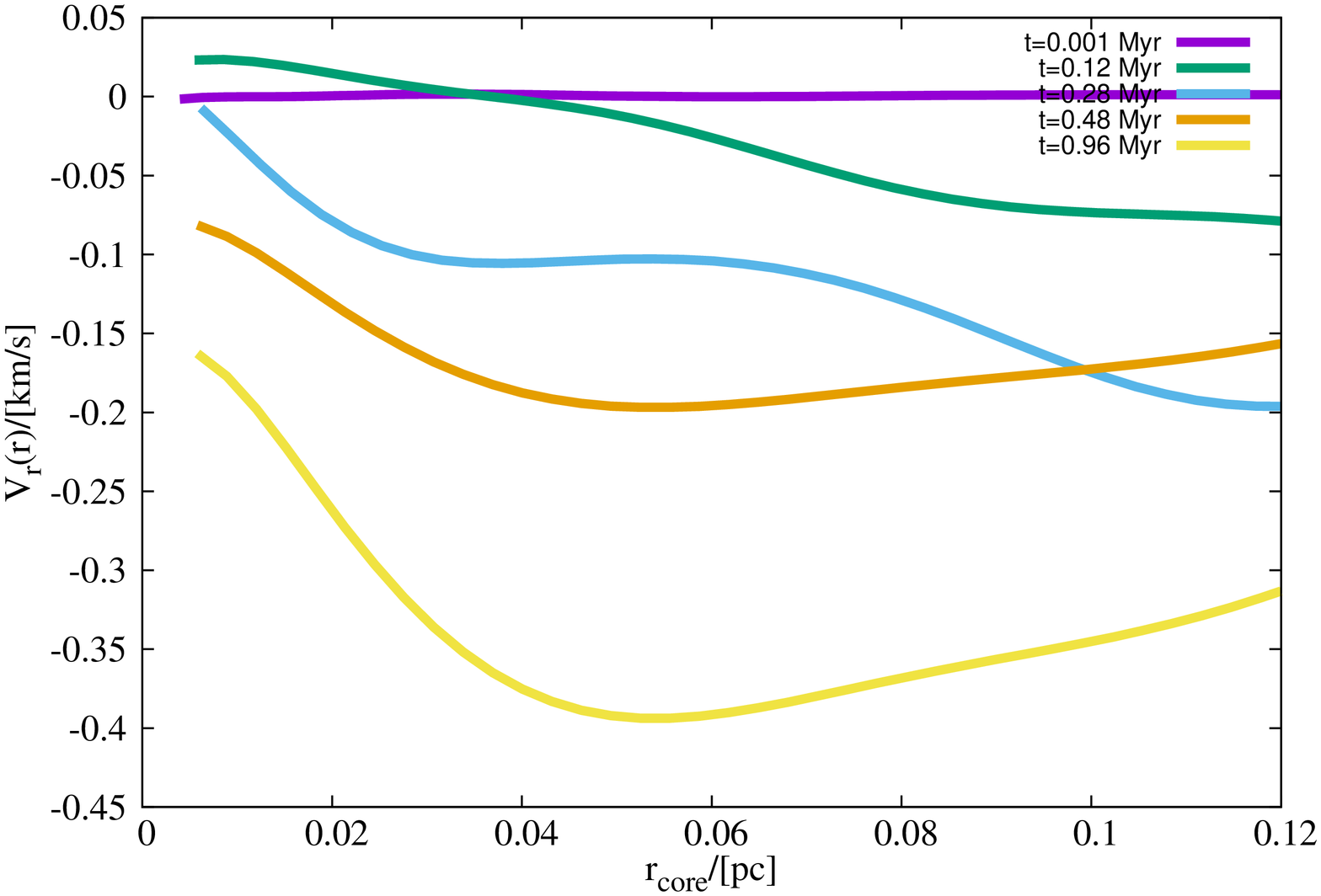}
\includegraphics[angle=270,width=0.5\textwidth]{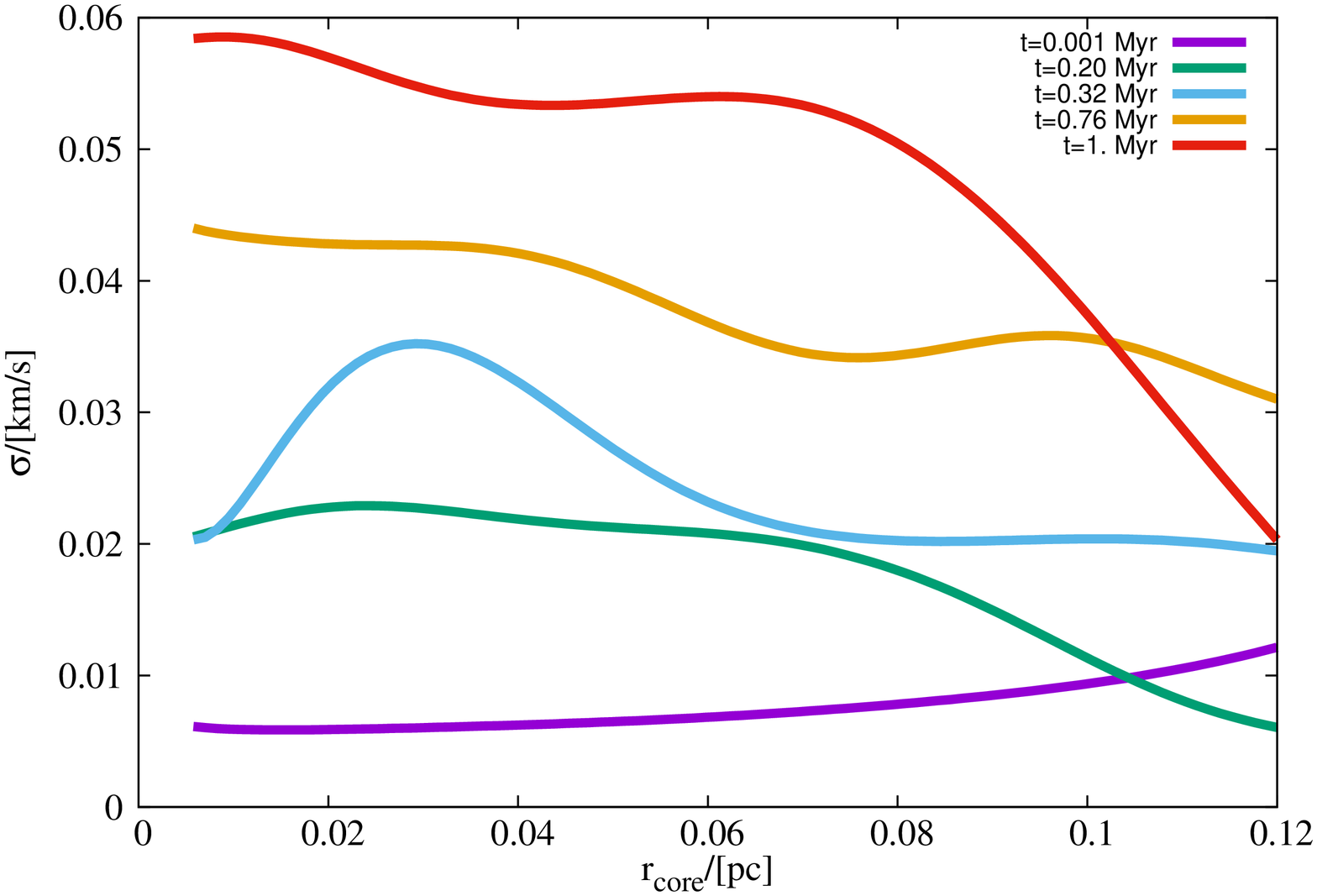}
\caption{\emph{Upper-panel :} As in the upper-panel of Fig. 5, but now the variation of the radial component of gas velocity in the core B68. \emph{Central-panel :} Radial component of the velocity of in-falling gas at different epochs of the evolution of the core L1521F(VeLLO), listed 8 in Table 1. \emph{Lower-panel :} Radial variation of gas-velocity dispersion at different epochs of the core B68 listed 1 in Table 1; \emph{see text for details.}}
\end{figure}
 Next, shown on the central-panel of this figure is the radial velocity of the in falling gas in the {\small VeLLO} {\small L1521F}, the magnitude of which is consistent with that reported by Onishi \emph{et al.} (1999). We note that the velocity of the in falling  gas in the envelope of the core in this case is transonic or even mildly supersonic in contrast to the gas-velocity observed for starless cores such as the {\small B68}. 
This inward/outward gas-motion in the core generates a velocity dispersion, the magnitude of which at different radii within the core {\small B68} has been shown on the lower-panel of Fig. 9.
Observe that the magnitude of velocity dispersion is relatively large close to the centre of the test core and its magnitude steadily increases as the core continues to evolve. Nevertheless,
the velocity field generated appears sufficient to arrest the collapse of the core. We will revisit this point in \S 3.5 with reference to the Bonnor Ebert stability criterion.
\subsection{Temperature profile within test cores}
Shown on the left-hand panel of Fig. 10 is the radial distribution of gas-temperature calculated using the standard grain-size in Eqn. (7) within the test cores at the epoch when the respective cores acquired their first density-peak. The radial temperature plots shown here were constructed by dividing the core into concentric shells followed by the calculation of the density averaged temperature for each shell. From this plot it is readily visible that although gas in the outer envelope of each of these cores remained relatively warm and approximately isothermal, gas-temperature close to their respective centre exhibited a sharp fall. It is only in this very small radius about the centre of the core that there is any evidence of significant cooling. Nevertheless, it proved insufficient to sustain collapse. However, the temperature profile and the magnitude of temperature deduced here for cores known to be starless viz., {\small B68}, {\small L694}, {\small L1517B} and {\small L1689-SMM2}, is consistent with that reported for these cores following several detailed observational studies discussed in \S 3.1 above. The core {\small L152F}, known to be a {\small VeLLO}, remained starless in our realisation(listed 7 in Table 1); where the standard dust grain-size was used in these calculations. \\ \\
Shown on the right-hand panel of Fig. 10 is the radial distribution of gas temperature within these cores early in their respective evolutionary cycle, but with the temperature now calculated using the puffed-up dust-grains; see \S 2.3. In this case, not only could we induce the core {\small L1521F} to collapse (see central-panel of Fig. 7), but all the other cores that previously remained starless also collapsed (see lower-panel of Fig. 7). The remarkable difference between the  plots on the respective panels of Fig. 10 probably hold the key to the question about the stability of a core, and its propensity to become protostellar. In this case gas closer to the centre of a core and therefore more dense than that in its envelope, appears to have cooled much more uniformly. This likely assisted the in-fall. \\ \\
Our simulations have shown that not all starless cores remain thermally sub-critical, i.e. $n_{c}\lesssim 10^{5}$ cm$^{-3}$. It appears that while a thermally super-critical core could be more likely to become protostellar, though it need not necessarily become protostellar. Consequently, a competing cooling mechanism must assist the gas to cool as a core continues to contract. This can be easily demonstrated with a simple deduction that follows. A core will become protostellar only if gas in it satisfies the Jeans criterion at all radii (Anathpindika \& Di Francesco 2013). Thus, $M(r) > M_{Jeans}(r)$, i.e.
\begin{equation}
\frac{4}{3}\Big(\frac{G^{3}}{\pi}\Big)^{1/2}r^{3}\rho^{3/2}(r) > a^{3}(r),
\end{equation}
where symbols have their usual meanings. Rearranging this equation yields,
\begin{displaymath}
\rho^{3/2}(r) > \frac{3}{4}\Big(\frac{\pi}{G^{3}}\Big)^{1/2}\frac{a^{3}}{r^{3}}.
\end{displaymath}
This equation yields the condition that gas density in a core must satisfy in order to become protostellar -
\begin{displaymath}
\frac{d\rho}{dr} > \frac{2\rho}{a}\Big(\frac{da}{r} - \frac{3a}{r}\Big),
\end{displaymath}
or equivalently, the sound-speed in the core must vary as
\begin{displaymath}
\frac{1}{a}\frac{da}{dr} < \frac{3}{r} + \frac{1}{2\rho}\frac{d\rho}{dr}.
\end{displaymath}
A simple integration of the above expression gives us the expression for sound-speed in a core that is likely to collapse -
\begin{equation}
a(r) < \mathcal{C}r^{3}\rho^{1/2},
\end{equation}
$\mathcal{C}$, being the constant of integration. If we denote the subscripts '1' and '2' to  denote the gas temperature on respectively the inward(toward core-centre) and outward (away from core-centre) side of $r$, then, $a(r)\equiv a_{1} - a_{2}$, moving out from the centre of the core. From Eqn. (19), we have
\begin{displaymath}
a_{1} < a_{2} + \mathcal{C}r^{2}\rho^{1/2},
\end{displaymath}
or after plugging the constant of integration
\begin{equation}
a_{1} < a_{2} + G^{1/2}\frac{r^{3}}{R_{core}^{2}}\rho^{1/2}(r).
\end{equation}
In other words, the sound-speed(or equivalently the temperature), looking inward to the centre of a core should decrease as defined by Eqn. (20) for collapse to ensue. \\ \\
\begin{figure*}
\vspace{1pc}
\mbox{\includegraphics[angle=270,width=0.5\textwidth]{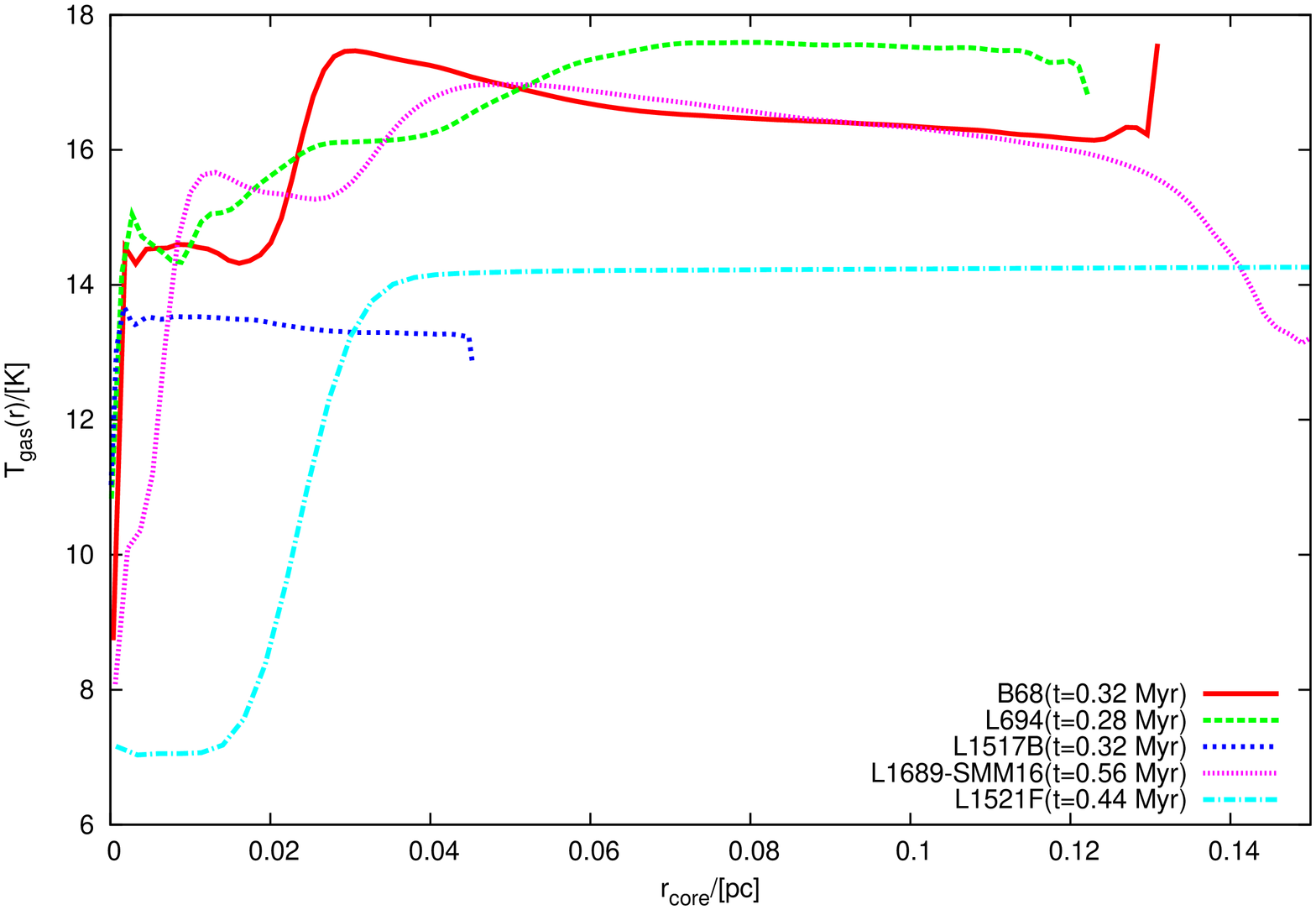}
\includegraphics[angle=270,width=0.5\textwidth]{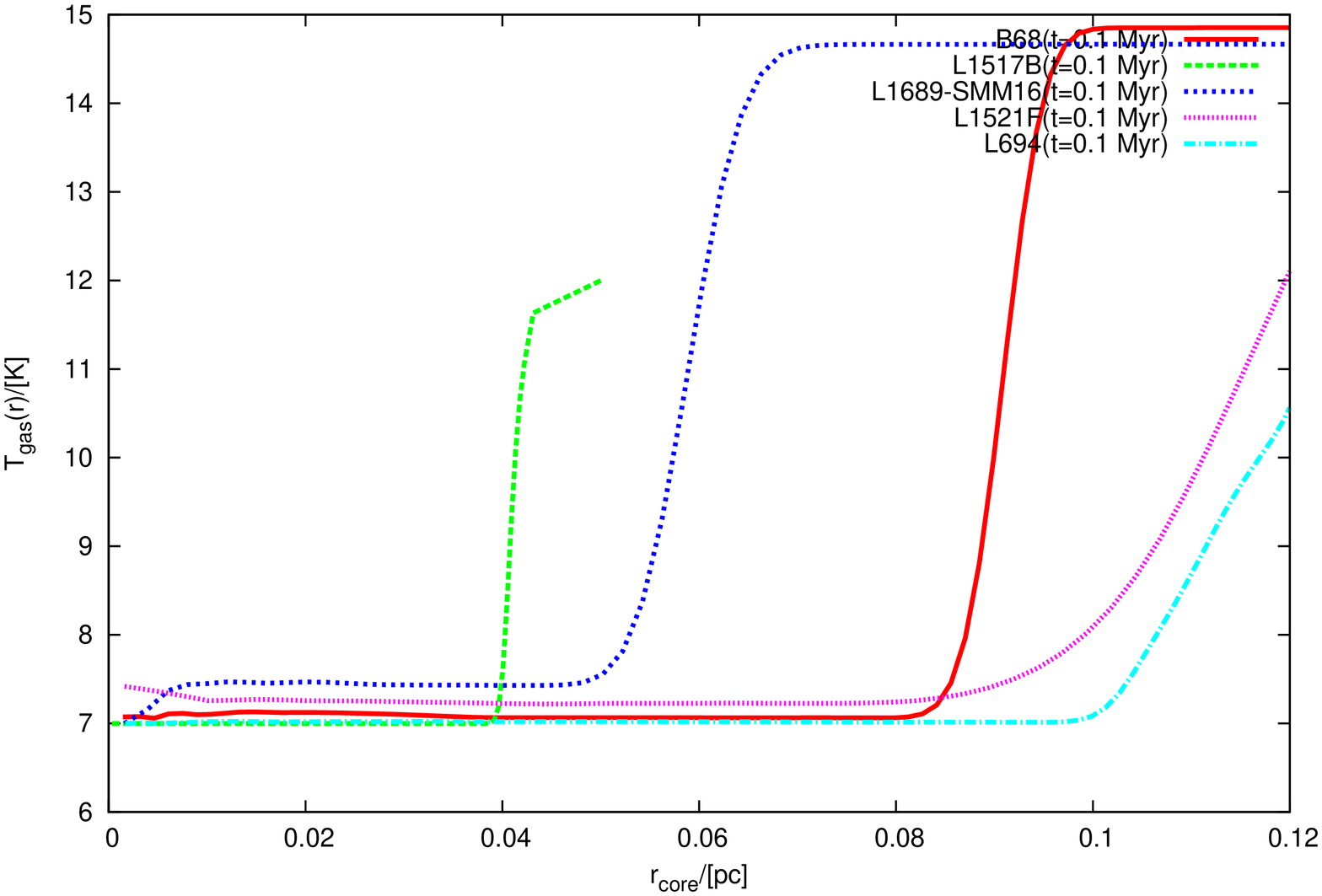}}
\caption{\emph{Left-panel} : As in Fig. 9, but now the radial variation of gas-temperature within respective test cores that remained starless at the epoch when they acquired their peak density; \emph{see text in \S 3.4}. \emph{Right-panel} : Radial distribution of gas temperature within respective cores with puffed-up dust-grains. Attention is especially drawn to the fact that in these latter realisations the cores developed a more extended cold region close to their centre fairly early in their evolutionary cycle.}
\end{figure*}

\subsection{Bonnor Ebert(BE) stability criterion}
Next, we undertake a comparison of the mass of the core, $M_{core}$, with its critical Bonnor-Ebert mass, $M_{BE}$, given by Eqn. (21) below, to ascertain if the Bonnor-Ebert mass is a reliable indicator of the virial boundedness of a core; 
\begin{equation}
M_{BE} = \frac{a_{0}^{4}}{(4\pi G^{3})^{1/2}P_{ext}^{1/2}}\cdot \mu(\xi_{b})\mathrm{e}^{-\psi(\xi_{b})},
\end{equation}  
$\mu(\xi)=\xi^{2}\frac{d\psi}{d\xi}$. For each of our test cores that remained starless (see Figs. 5, 6 \& 7), we used the respective magnitude of $\xi_{b}$, the radius of the {\small BES} whose density profile fitted the observed density distribution of our model cores, to calculate $M_{BE}$. For instance, in the specific case of the core {\small B68}, the above equation simply becomes,
\begin{equation}
M_{BE} = 1.125\frac{a_{0}^{4}}{G^{3/2}P_{ext}^{1/2}}.
\end{equation}
The upshot of plugging in the appropriate values of gas temperature, $\sim$ 15 K for the core {\small B68}(see left-hand panel of Fig. 10; $t\sim$ 0.32 Myr), and $\frac{P_{ext}}{k_{B}}\sim 4.5\times 10^{4}$ K cm$^{-3}$, after including velocity dispersion of the {\small ICM}, is, $M_{BE}\sim$ 2.66 M$_{\odot}$; $\frac{M_{core}}{M_{BE}} < $1. In other words, the core cannot possibly become singular, which in fact, is consistent with the observed fate of the model core. \\ \\
Similarly, plugging-in the sound-speed corresponding to the gas temperature ($\sim$ 16 K, $\sim$ 13 K and $\sim$16 K for respectively the cores {\small L694-2}, {\small L1517B} and {\small L1689-SMM2}), along with the  radius of the {\small BES} whose profile fitted the extant density distribution for the individual core (see plots in Fig. 6), we find that  $\frac{M_{core}}{M_{BE}} >$1 for each of the three cores. Evidently, the comparison between the {\small BE} mass and the core mass yields contradicting inferences as regards the virial stability of a core, for the unstable {\small BES}, by definition, is susceptible to collapse. However, as discussed above, such a collapse did not materialise in any of these cores leading us to conclude, the {\small BE} stability criterion is a unreliable predictor of the future evolution of a test core. The problem with this analysis is alleviated when the sound-speed in Eqn. (21) is corrected to include the velocity dispersion (see e.g., Johnstone \emph{et al.} 2000). Our inference is consistent with the recent findings reported by Pattle et al. (2015) where a number of cores surveyed in the Ophiuchus as part of the {\small JCMT} Gould Belt survey\footnote{http://www.jach.hawaii.edu/JCMT/surveys/}, though unstable according to the Bonnor-Ebert stability criterion, did not show any credible signs of star-formation.
\begin{figure*}
\vspace{1pc}
\mbox{\includegraphics[angle=0,width=0.33\textwidth]{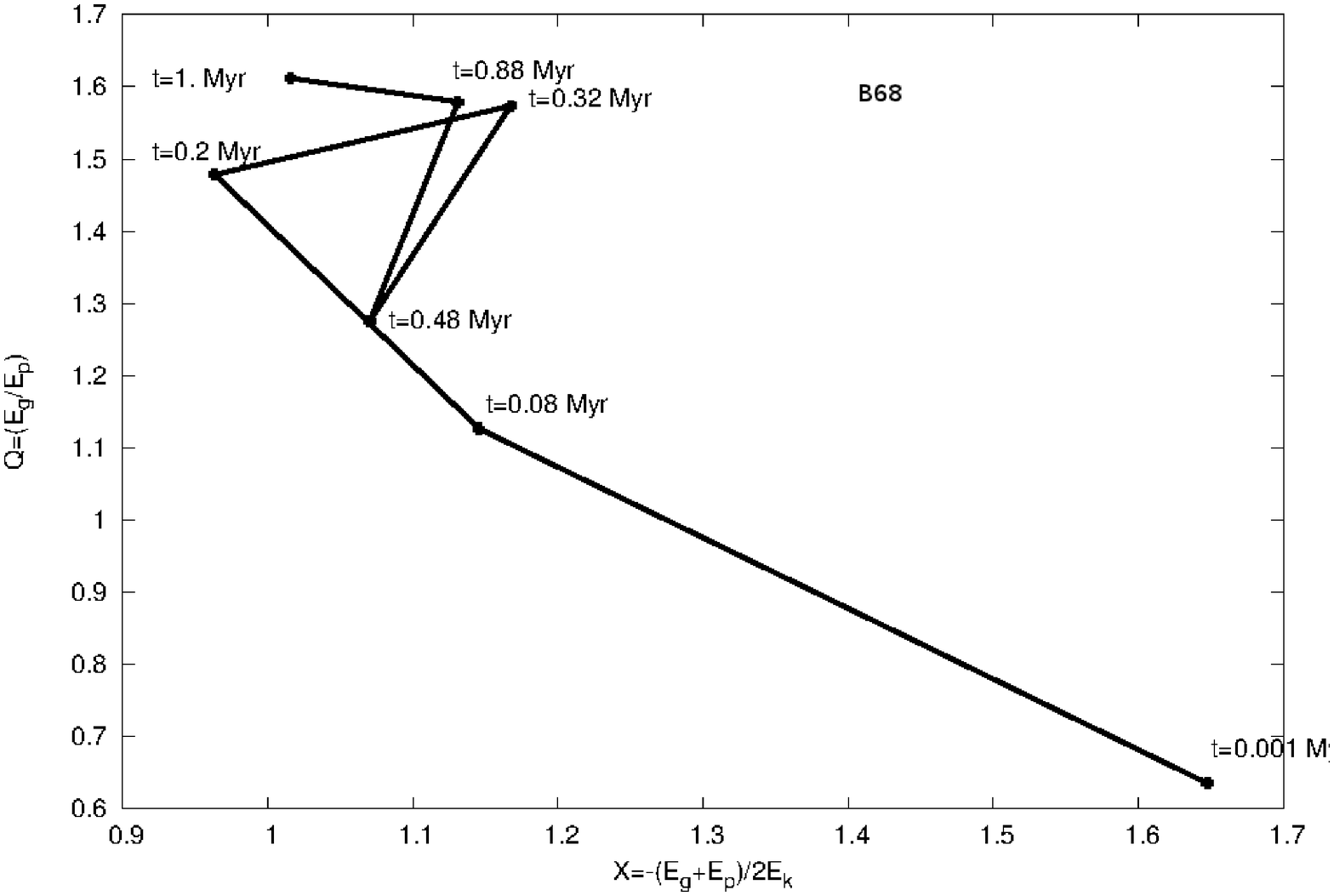}
\includegraphics[angle=0,width=0.33\textwidth]{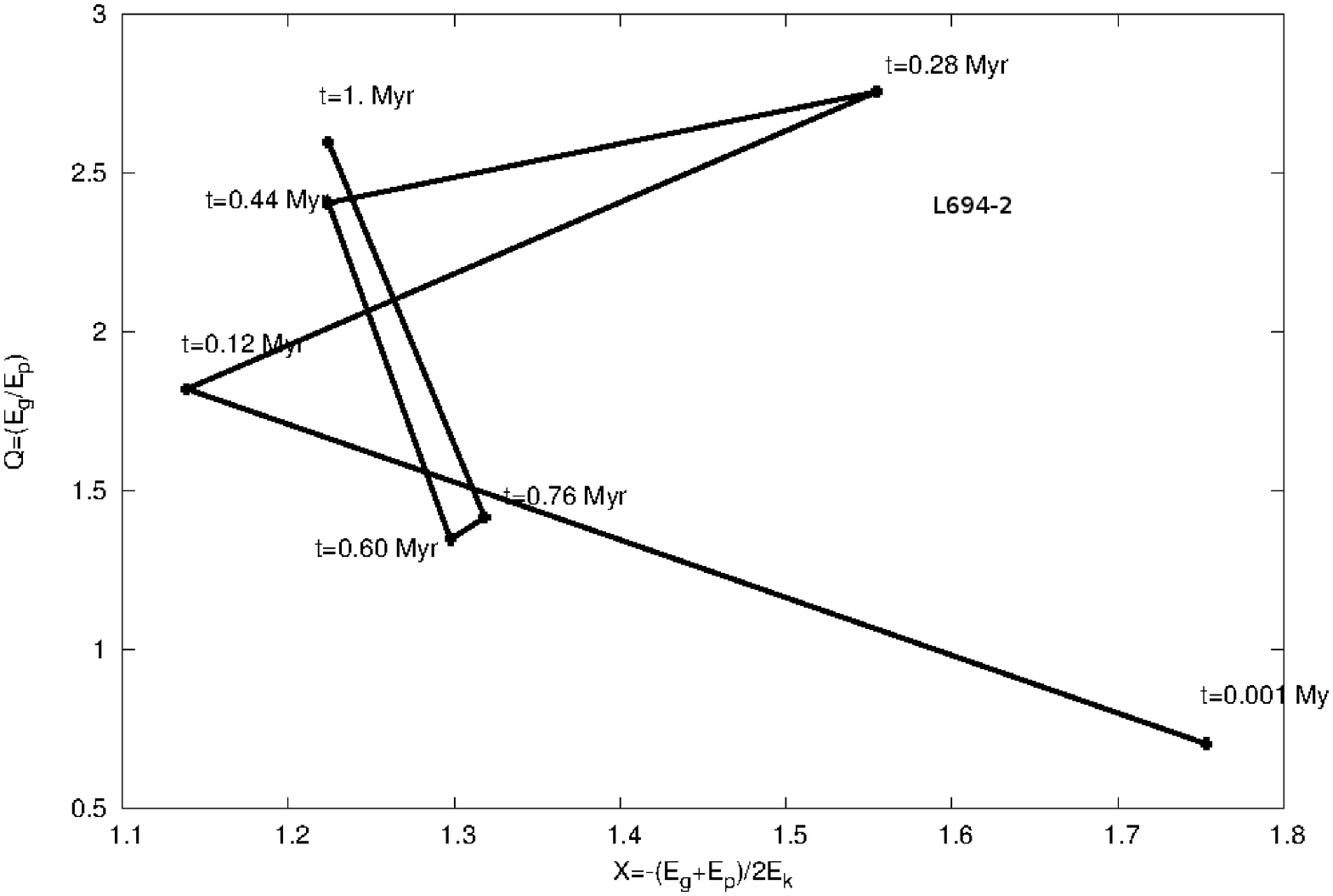}
\includegraphics[angle=0,width=0.33\textwidth]{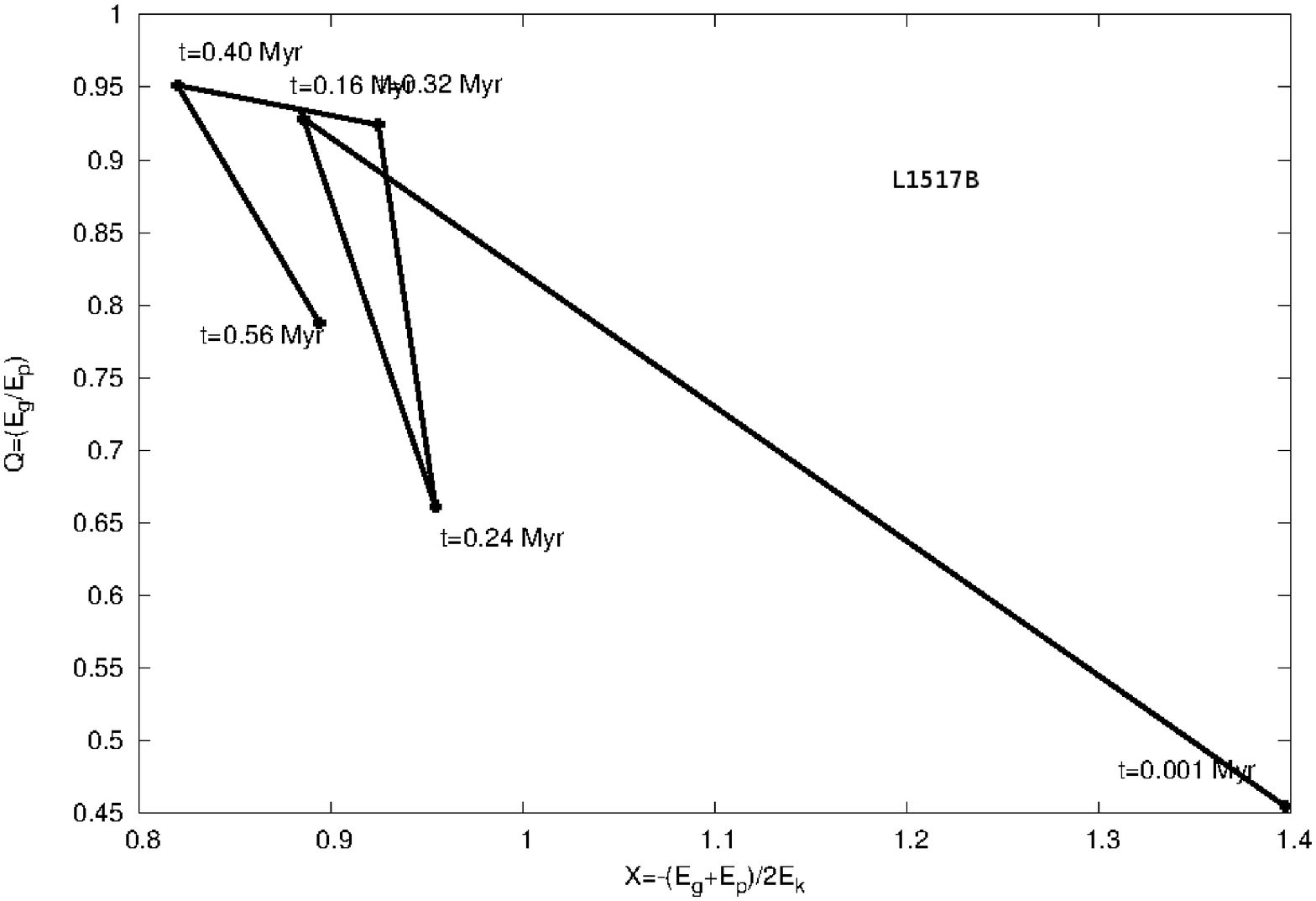}}
\mbox{\includegraphics[angle=0,width=0.33\textwidth]{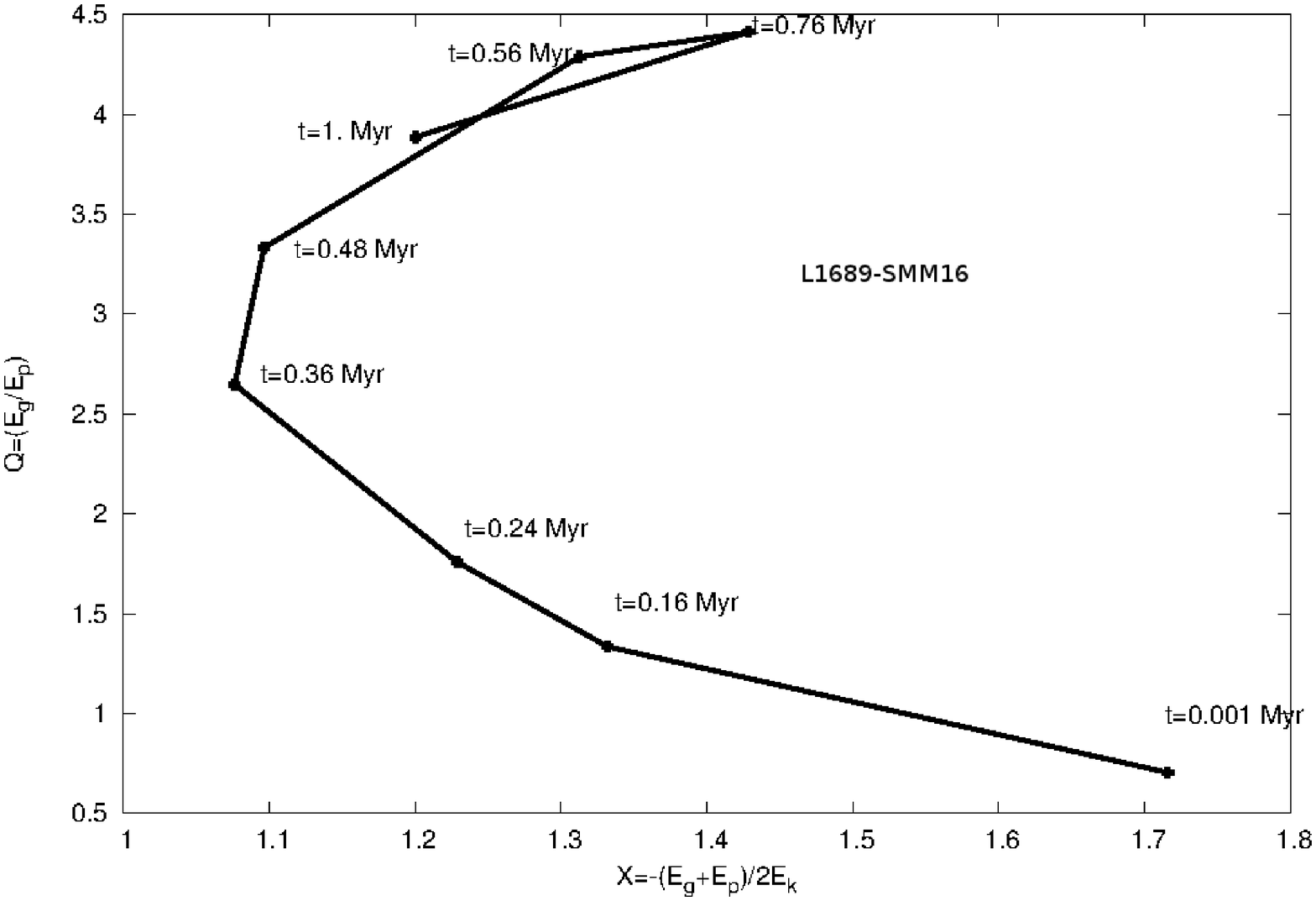}
\includegraphics[angle=0,width=0.33\textwidth]{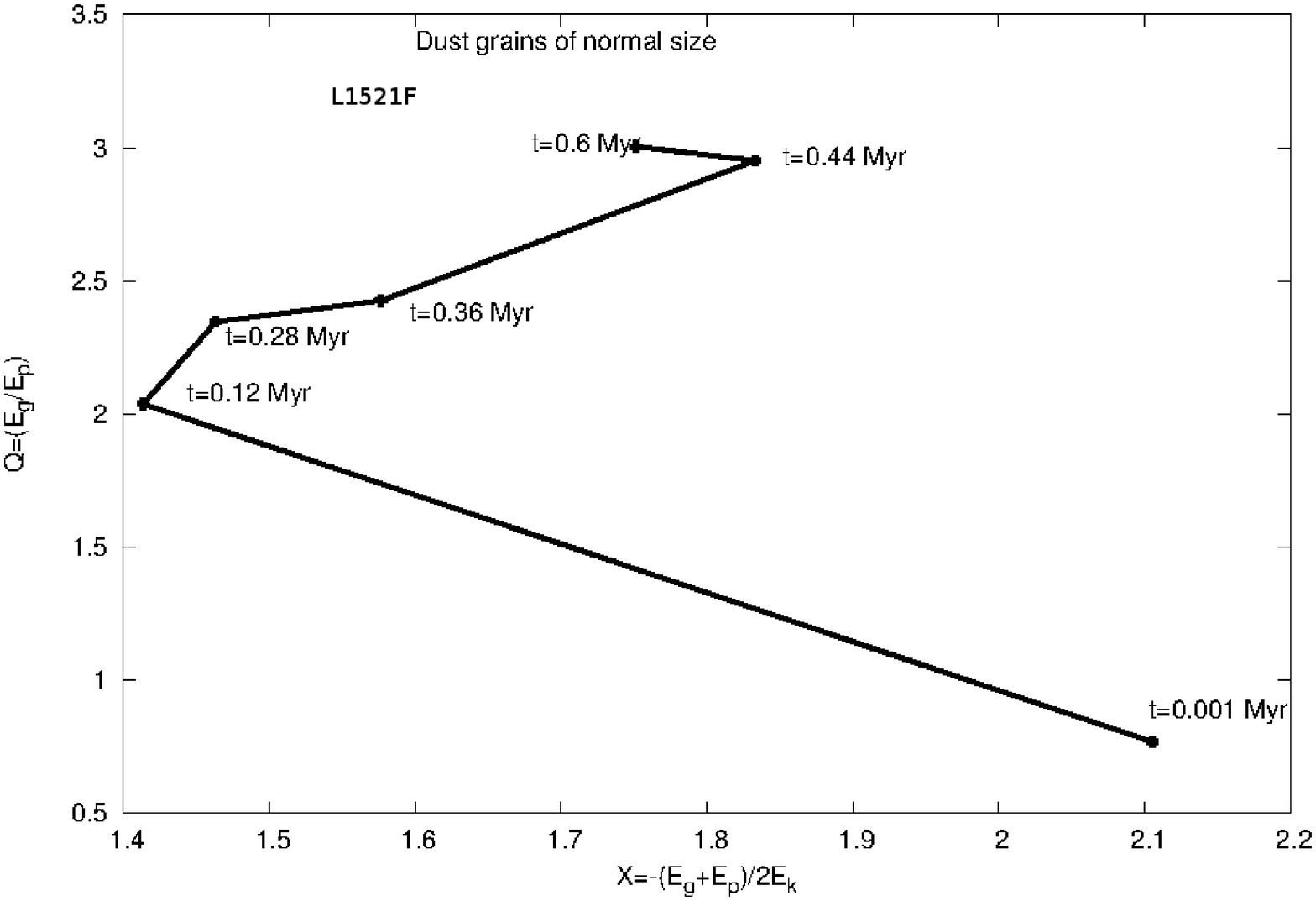}
\includegraphics[angle=0,width=0.33\textwidth]{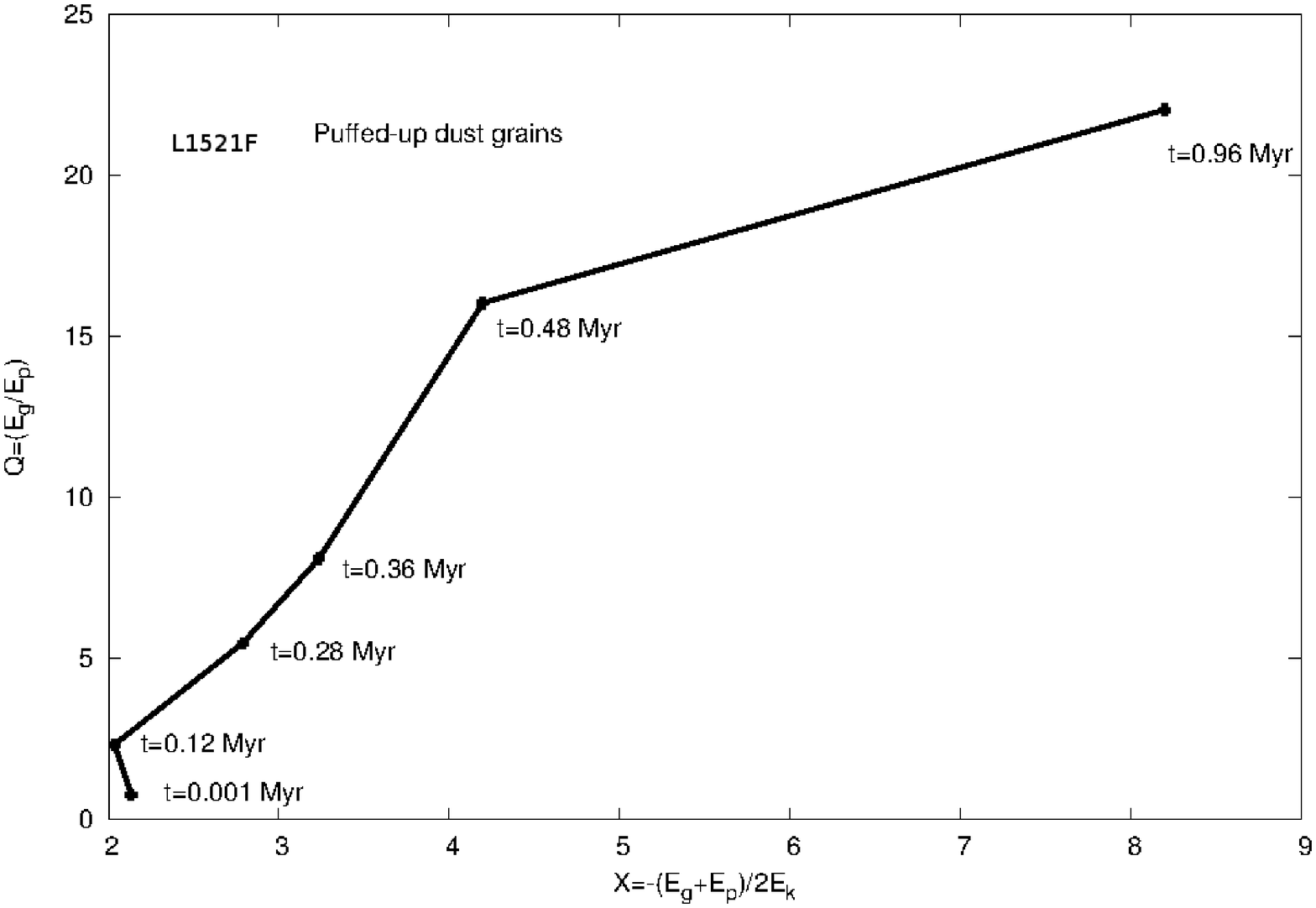}}
\caption{Plots showing the temporal excursion of test cores on their respective Viral chart. The region right-ward of $X >$1 corresponds to the virially bound state where-as to the left, virially unbound. Similarly, the region upward of $Q >$1, corresponds to the state when a core is dominated by self-gravity, while that below, to one that is confined by external-pressure.}
\end{figure*}
\subsection{The Virial chart for cores}
Finally, we discuss the virial state of our test cores over the course of their evolution. For a purely hydrodynamic calculation the virial equation is of the form -
\begin{equation}
\frac{1}{2}\ddot{\mathrm{I}} = 2E_{k} + E_{g} + E_{p},
\end{equation}
$E_{k}, E_{g}$ and $E_{p}$ being respectively the kinetic energy, gravitational energy and the energy due to the pressure confining the core. The second moment of inertia, $\ddot{\mathrm{I}}$, is greater than zero for a dispersing core, less than zero for one that is in-falling and equal to zero for a core in virial equilibrium. The virial components for the cores tested in this work were calculated as follows -
\begin{equation}
E_{k} = \frac{3}{2}\Sigma_{i}m_{i}\sigma_{i}^{2},
\end{equation}
where the summation extends over all the gas particles and $\sigma_{i} = (a_{i} + (\sigma_{v})_{i})$; $a_{i}$, being the sound-speed for a particle, $(\sigma_{v})_{i}$, its velocity dispersion and $m_{i}$, the mass of individual particle. The gravitational energy, $E_{g}$, as
\begin{equation}
E_{g} = -GM_{core}\Sigma_{i}\frac{m_{i}}{\vert \mathbf{r}_{c} - \mathbf{r}_{i}\vert}.
\end{equation}
As in Eqn. (24) the summation here also runs over all gas particles; $\mathbf{r}_{c}$, is coordinate of the centre of the core and, $\mathbf{r}_{i}$, the position of an individual particle in the core. Finally, the energy corresponding to the magnitude of external pressure acting on the core
\begin{equation}
E_{p} = P_{ext}V_{core}\equiv k_{B}V_{core}\bar{n}_{icm}T_{icm},
\end{equation}
where $V_{core}$ is the volume of the test-core and, $\bar{n}_{icm}$, the mean particle density of the externally confining medium. We now define two quantities, $Q\equiv\frac{E_{g}}{E_{p}}$, that quantifies the strength of self-gravity relative to the external pressure, and $X\equiv\frac{-(E_{g} + E_{p})}{2E_{k}}$, the virial ratio. Note that the energy, $E_{p}$, exerted by the externally confining medium, like the gravitational energy, $E_{g}$, is also taken as negative since it is inwardly directed. \\ \\ 
A core is defined to be virially bound when $X\geq 1$, and pressure-confined, when $Q < 1$ (e.g. Pattle \emph{et al.} 2015). Shown in Fig. 11 is the variation in the magnitude of these quantities at different epochs of our test cores; note that the Virial chart for {\small B68} shown here is that for the core modelled as a stable {\small BES}($\xi_{b}$=3).
As is visible from the respective charts for our model cores, each of these cores started as a virially bound and pressure-confined entity($t\sim$0.001 Myr for each). As the cores continued to contract and acquire a centrally concentrated density-profile, then on the one hand the extent of gravitational boundedness in each of these cores can be seen to be increasing steadily, but on the other, they tended to become less virially bound.  In the latter stages of their evolution, however, cores that remained starless slided down the chart and eventually acquired a configuration that is both, gravitationally bound ($Q > $1), as well as virially bound. It appears, under normal circumstances, these cores will remain starless, albeit bounded. 
The Virial-chart on the lower right-hand panel for the core {\small L1521F}, the {\small VeLLO}, where the gas-temperature was calculated with puffed-up dust-grains is significantly different from that for the starless cores; again in the interest of brevity we have shown here the virial chart for only this core although calculations were repeated with dust-grains of enhanced cross-section for all the cores. Indeed, the extent of gravitational boundedness is much higher for the {\small VeLLO} as reflected by a significantly larger magnitude of $Q$, in comparison with that for the {\small L1521F} that rebounded. Furthermore, after its rebound, this latter core acquired a state of approximate equilibrium where the self-gravity was relatively stronger than the external-pressure. On the other hand, self-gravity continued to dominate the {\small VeLLO}. In the next section we will discuss the implications of this observation on the ability of a core to form stars. 

\section{Discussion}
\subsection{Evolution of prestellar cores}
A \emph{unified picture} of core-formation and evolution in which a putative star-forming core is assembled in the outside-in fashion by converging gas-flows, has recently been suggested by Gong \& Ostriker (2009, 2011). The assembled core in this case remains stable as long as its mass is less than the critical mass, i.e., the critical mass, $M_{BE}$, for the marginally stable {\small BES}. Once the core acquires mass greater than $M_{BE}$, it begins to collapse in the inside-out fashion. There are a few difficulties with this proposition : \textbf{(i)} recent observations of several starless cores have shown that despite individual cores having a density profile resembling that of a unstable {\small BES}, they do not show any discernible signs of collapse (e.g. {\small B68}, Alves \emph{et al.} 2001), \textbf{(ii)} it is difficult to perceive how nature could possibly assemble a core that is marginally stable. Furthermore, recent observations of a number of starless cores have also shown that some cores with masses exceeding their thermal Jeans mass  or indeed, their critical {BE} mass do in fact, remain starless, the so-called \emph{Super Jeans} cores (e.g. Sadavoy \emph{et al.} 2010 a,b), \textbf{(iii)} if nature does indeed prefer to assemble cores via converging supersonic gas-flows, in that case the broadening of spectral emission lines from cores should be significantly broader than observed to reflect the agitated nature of gas within a core, and finally \textbf{(iv)} the inside-out collapse is inconsistent with observations of a number of starless cores that exhibit inwardly directed gas-flow (e.g. Tafalla \emph{et al.} 1998, 2004; Lee \emph{et al.}2007, Schnee \emph{et al.} 2007), with velocity magnitude on the order of $\lesssim$0.1 km/s which is significantly smaller than the typical free-fall velocity. \\ \\
In the present work we modelled each of our test cores as a pressure-confined {\small BES} ($\xi_{b}$=3), which was then allowed to evolve by itself in the presence of self-gravity. The gas temperature was calculated in a self-consistent manner by explicitly solving the respective equations of thermal balance for dust and gas.  We demonstrated that the eventual fate befalling a core bore largely on the temperature profile it developed rather than on its initial density distribution. In the specific case of the core {\small B68} we showed that merely assembling the test core as a unstable {\small BES}  was insufficient to induce it to collapse to singularity. We observed, irrespective of the choice of the polytrope used to model this core, with the standard dust grain-size, the test core did not collapse even after acquiring a density distribution that mimicked the profile of a unstable {\small BES}. In this instance although our test cores developed a cold central region with gas temperature $\lesssim$10 K, they remained starless. In fact we argue that a core can possibly become singular only if gas in its interiors can cool efficiently during its contraction so as to be able to sustain its collapse. \\ \\
To establish our argument about the importance of gas temperature in the evolution of a prestellar core, we performed two realisations for the test core {\small L1521F}. This core is believed to have formed  its first hydrostatic core and has been classified as a {\small VeLLO} in literature (e.g. Takahashi \emph{et al.} 2013). In the realisation listed 7 in Table 1, performed using the standard grain-size for dust in the interstellar medium, the core was seen to rebound. The core in the realisation listed 8 in Table 1 where a higher dust grain-size was adopted, collapsed after the gas close to its centre acquired a temperature between $\sim$7 K and 10 K. We repeated calculations for all the other starless cores viz., {\small B68, L694-2, L1517B} and {\small L1689-SMM2} with this higher grain-size. It was observed that the test cores in all these latter realisations developed cold interiors and collapsed. Evidently, the temperature gradient in a core or in other words, the extent of cooling within a core  must somehow determine if whether it will remain starless or become protostellar. This inference begs the question about the possible mechanism/s that might assist gas cooling. It is well-known that gas in a core cools primarily via ro-vibrational transitions of excited molecules and through the interaction between gas and dust, provided of course that the dust is cooler than gas. Detailed modelling of the cooling function for molecular gas in dark clouds by Goldsmith (2001) has shown that the molecular cooling becomes significantly stronger for gas densities upward of $\sim 10^{5}$ cm$^{-3}$, a phase described as thermally super-critical in literature (e.g. Keto \& Caselli 2008). This remains true even after accounting for the depletion of molecular species at higher densities. \\ \\
\begin{table}
 \centering
 \begin{minipage}{0.5\textwidth}
  \caption{The observed lifetime of model cores.}
  \begin{tabular}{@{}llll}
  \hline
   Core & Peak density of & $t_{ff}$ & $t_{c}$\\
   Name & the model core  & [Myrs]\footnote{Free-fall time  corresponding to the peak density acquired by the core.} & [Myrs]\footnote{Contraction timescale for the core (\emph{see text}).}\\
        & [$\times 10^{4}$ cm$^{-3}$] & & \\
  \hline
   B68($\xi_{b}$=3) & 3.5 & 0.17 & 0.32 \\
   B68($\xi_{b}$=6.5)& 30.0 &0.07 & 0.36\\
   B68(Uniform density)&10&0.11 & 0.16\\
   L694-2 & 15.0 & 0.09 & 0.28 \\
   L1517B & 4.0 & 0.17 & 0.32 \\
   L1689  & 25.0 & 0.07 & 0.56 \\
   L1521F\footnote{Listed 7 in Table 1}& 20. & 0.08 & 0.44 \\
  \hline     
  \end{tabular}
 \end{minipage}
\end{table}
Next, a cursory glance at Column 2 in Table 2 shows that some of our test cores remained starless despite becoming thermally super-critical. Earlier in \S 3 we have already ruled out the critically stable {\small BES} and the uniform density sphere as a model core for the B68 so the corresponding entries need not be considered here. \emph{It is however, quite clear from this table that acquiring the thermally super-critical state for a core is not sufficient to ensure its collapse. }
\subsection{Prestellar core lifetime} 
Listed in column 3 of Table 2 is the contraction timescale($t_{c}$), i.e., the timescale over which a core acquires its peak density, for our test cores. We note, this timescale is on the order of $\sim 10^{5}$ years for each test core. Thus, the model cores in this work appear to have contracted in a quasistatic manner over a few free-fall times and easily survived much longer (the simulations were terminated much later after the cores had contracted).
Let us now examine the observational evidence on this count. In one of the early surveys of prestellar cores, Lee \emph{et al.}(1999) found 220 cores of which 17($\sim$ 8\%) showed signs of in-fall; specifically, 7 cores definitely showed evidence of in-fall while 10 others showed tentative evidence of in-fall. Later,Gregersen \& Evans (2000) surveyed a subset of about 50 cores originally identified by Beichman \emph{et al.}(1986). This survey identified 6 cores ($\sim$ 10\%) with signs of in-fall. Similarly, Enoch \emph{et al.}(2007) identified 92 protostellar objects out of 201 cores surveyed collectively in  the Ophiuchus, Perseus and the Serpens  molecular clouds. In other words, $\sim$ 46\% of the cores in this sample had collapsed. More recently, Schnee \emph{et al.}(2013) reported observations of 26 starless cores in the Perseus molecular cloud, 11($\sim$42 \%) of which showed signs of in-fall. \\ \\
Now, observationally the total lifetime of starless cores is estimated from the number ratio of cores with and without a protostar. This estimate is on the order of, $t_{SL}\sim$0.3 - 1.6 Myrs (Lee \& Myers 1999), which is consistent with earlier estimates suggested by Beichman \emph{et al. 1986} and Ward-Thompson \emph{et al.}(1994); see also Ward-Thompson \emph{et al.}(2007). Using the fraction, $f_{inf}$, of cores showing signs of in-fall from literature the typical contraction timescale, $t_{c}$, of a prestellar core can be approximated as, $t_{c}\sim t_{SL}*f_{inf}$, estimates of which vary from $8\times 10^{4}$ years for data from Lee \emph{et al.}(1999), $\sim 10^{5}$ years for that from Gregersen \& Evans (2000), $\sim 5\times 10^{5}$ years for data from Enoch \emph{et al.}(2007), and $\sim 4\times 10^{5}$ years for data from Schnee \emph{et al.}(2013). Evidently, the estimate of $t_{c}$ using the data from Lee \emph{et al.}(1999) is somewhat asynchronous with the magnitude calculated from the rest which are mutually consistent. \\ \\
Variation in the number of cores showing signs of in-fall is therefore not totally surprising. On the other hand, the contraction timescale, $t_{c}$, observed for the test cores in this work is also consistent with the chemical age reported for some starless cores. Maret \emph{et al.}(2013), for instance, estimated the chemical age for cores {\small L1517B} and {L1498} on the order of a few 10$^{5}$ years by radiative transfer modelling of these respective cores, coupled with a detailed chemical network. \emph{Nevertheless, the timescale over which the model cores in this work contracted to acquire their respective peak density is consistent with the timescale, $t_{c}$, deduced from the surveys reported above. Equally significant is the fact that we could reconcile the observed longevity of starless cores with purely hydrodynamic models without including any additional support from the magnetic field and/or turbulent velocity field.} \\ \\

\begin{table}
 \centering
 \begin{minipage}{0.5\textwidth}
  \caption{virial coefficients for test cores(modelled as Bonnor-Ebert sphere with $\xi_{b}$=3) at the epoch when they reach their maximum density.}
  \begin{tabular}{@{}lll}
  \hline
   Core & Q & X \\
    Name & &  \\
  \hline
    B68 & 1.57 & 1.17  \\
    L694 & 2.75 & 1.55 \\
    L1517B & 0.92 & 0.93 \\
    L1689 & 4.29 & 1.31 \\
    L1521F(Rebounded) & 3.0 & 1.75 \\
    L1521F({\small VeLLO}; at $t\sim$0.36 Myr) & 8.11 & 3.24 \\
  \hline
\end{tabular}
 \end{minipage}
\end{table}

\subsection{Virial state of cores}
In a more recent contribution Pattle \emph{et al.}(2015) discussed the stability of a sample of cores detected in the Ophiuchus {\small MC} with the {\small SCUBA-2} instrument on the {\small JCMT}. They found that a majority of the cores in their sample were either gravitationally or virially bound. We introduce here the virial chart, a plot showing the temporal evolution of the virial components of the test core. The virial chart for our test cores show that over the course of its evolution, a core can acquire a configuration that is gravitationally bound, pressure-confined or virially bound. The rate at which a core evolves of course, depends on its temperature profile. Furthermore, as has already been pointed out by Pattle \emph{et al.}(2015), the Bonnor-Ebert mass, $M_{BE}$, is not a particularly good indicator of the propensity of a core to become protostellar. \\ \\
Simulations discussed in this work also demonstrate that the mere fulfilment of the criterion, $\frac{M_{core}}{M_{BE}}>$1, is not sufficient for a core to become singular and in fact, correction of the sound-speed in the expression for $M_{BE}$ to include the velocity dispersion appears to be a better test to determine the stability of a core.
The second and third columns in Table 3 call for attention in this regard. Listed in these columns are the virial coefficients for the respective cores at the epoch when they were centrally most concentrated. A cursory glance shows that the respective cores were both gravitationally as well as virially bound with {\small L1517B} being the only exception as it remained pressure-confined, 
as well as virially unbound. Consequently, all these starless cores(except perhaps {\small L1517B}), were good enough to become protostellar and yet, remained starless. This leads to the next conclusion : \emph{being gravitationally and/or virially bound, though necessary, is not a sufficient condition for a core to become protostellar. Realisations of the core {\small L1521F}, listed 7 and 8 in Table 1, are good examples to illustrate this point.}
\subsection{The case for enhanced dust grain-size}
The temperature of gas in a core is sensitive to the strength of the cosmic-ray({\small CR}) ionisation rate. 
Keto \& Caselli (2008) demonstrated that a {\small CR} ionisation rate stronger than average raised the gas temperature in the outer regions of a core irrespective of whether it was thermally sub-critical or super-critical. Furthermore, these authors also demonstrated that the temperature of gas close to the centre of a thermally super-critical core was lowered significantly to $\sim$6 K when the dust opacity was raised by a factor of four. This, however, did not appear to lower the gas temperature below $\sim$10 K close to the centre of a sub-critical core. Presently there is some tentative evidence to support the idea of enhanced opacity due to relatively larger, i.e., fluffy dust grains.
 One of the first suggestions in this regard was made by Kr$\ddot{u}$gel \& Siebenmorgen (1994) after comparing CO and dust observations for a few typical dark clouds. They argued, dust grains in dark clouds likely augmented their size as depleted molecular species gradually adsorbed on the grain surface. These dust grains could possibly coagulate to form the so-called puffed-up, fluffy dust grains with enhanced opacity probably then followed (e.g. Ossenkopf \& Henning 1994). Indirect evidence in favour of higher dust opacity has also been suggested on the basis of estimates of core masses. Evans \emph{et al.}(2001), for instance, derived core masses using the standard dust opacity and found them well in excess of the maximum stable mass for a {\small BES}. On the contrary, raising the dust opacity lowered the estimates of core mass such that they were then consistent with the Bonnor-Ebert stability criterion. Similarly, Keto \emph{et al.}(2004), who calculated core masses using the N$_{2}$H$^{+}$ lines found that opacities of dust grains would have to be significantly higher than that for the {\small ISM} in order to be consistent with the core masses calculated in the survey. \\ \\
The possibility of grain-growth can also be studied from the inferred values of the dust opacity spectral index that is known to be sensitive to the maximum grain-size, but insensitive to fine grains smaller than $\sim$1 $\mu$m. Detailed models that study the growth of dust-grains have shown that the dust opacity spectral index lies between $\sim$1.5-2 for fine granular dust, then increases to between 2-3 for grains typically on the order of $\sim$10$^{2}$$\mu$ m - $\sim$10$^{3}$$\mu$ m and then falls to less than unity for even bigger dust particles. By comparison, the opacity spectral index for dust in the interstellar medium is $\sim$2 (see e.g. Testi \emph{et al.} 2014). Interestingly, in a more recent work Chen \emph{et al.}(2015) have reported dust spectral indices between 2-3 and between 0-1.5 towards a few clumps in the Perseus {\small MC}. These magnitudes suggest that dust coagulation leading to the formation of larger grains is perhaps underway in these clumps. In a recent work that lends more support to this conclusion, Juvela et al. (2015) have reported a median dust spectral index of $\sim$1.84 for cores detected in the \emph{Galactic Cold cores} survey carried out with data from the Herschel and Planck. \\ \\
Another point that favours puffed-up dust grains is the fact that
the latter tends to lower gas temperature in a core more uniformly as is visible in the plot on the right-hand panel of Fig. 10 for {\small L1521F} and indeed for all the other test cores. The plot on the right-hand panel of this figure is more consistent with the temperature($\sim$6K-7K), found close to the centre of typical cores (e.g. Crapsi \emph{et al.} 2007), and indeed, toward the centre of the {\small VeLLO} {\small L1521F} (Chitsazzadeh 2014). Puffed-up dust grains have a higher opacity and enhance gas-cooling by efficiently coupling with the gas that effectively lowers the gas temperature within a core. 

\subsection{Other similar work and limitations of this work} The proposition of modelling a prestellar core with a pressure-confined {\small BES} is qualitatively similar to a number of models suggested in the past. Like those models, our model also has a central region that has approximately uniform density, beyond which it peters off. Such models have been discussed by Henriksen \emph{et al.}(1997) and Whitworth \& Ward-Thompson (2001). While Henriksen \emph{et al.}(1997) assumed a constant central density, Whitworth \& Ward-Thompson (2001) assumed the central density to be rising according to a power-law. Models using ambipolar diffusion have predicted similar density profiles (e.g. Safier \emph{et al.}1997). The model core in this latter case contracted on a timescale longer than popular estimates of contraction time, $t_{c}$, and could not reproduce the observed relatively large in-fall velocities. On the other hand, we present a more general model for the initial density distribution of a typical prestellar core. Our model is inspired by the profile of a core extracted from a previously developed numerical simulation aimed at studying the formation of prestellar cores. It does not have any free parameters.  \emph{Importantly, it is not our endeavour to fit either the density distribution or the distribution of gas temperature within a core at any particular epoch of its evolution. Instead, we follow the temporal evolution of a test core in the presence of self-gravity. By doing so, we derived the density distribution, the temperature profile and the velocity of in-falling gas at different epochs of evolution of a test core.  In general, results from our realisations  are consistent with the findings reported by various authors in literature for the respective cores. }
 \\ \\
\textbf{Limitations} Without a full-fledged chemical network at our disposal, gas temperature in realisations discussed here was calculated by employing various cooling functions to account for heating/cooling of gas and dust. Consequently, the calculations presented here do not account for depletion of various molecular species at higher gas densities. As a result, the cooling-rate at higher densities is likely to have been somewhat over-estimated in these calculations. That, however, is unlikely to significantly affect our conclusions, for the cooling-function due to depleted molecular species behaves approximately similarly to the function for undepleted species, though the cooling-rate is somewhat lower leading to slightly higher(by only a fraction of a Kelvin as a result of CO depletion) gas temperature (Goldsmith 2001; Keto \& Caselli 2008). This is simply because the equilibration timescale of the CO is only a fraction of the dynamical timescale of a core; in other words, the depleted CO is rapidly replenished via the inverse process of desorption (Keto \& Caselli 2010). Furthermore, employment of a formal radiative-transfer scheme to calculate gas and dust temperature would have permitted us to asses the extent to which the {\small ISRF} is attenuated as a function of the column density of the test core. In the present work though, we have assumed constant heating due to the {\small ISRF} with a fixed attenuation within a test-core and characterised by the parameter, $\chi\sim 10^{-4}$. This is a reasonable assumption, consistent with magnitudes reported for typical {\small IRDC}s(e.g. Goldsmith 2001; Valdivia et al. 2016). We also note that conclusions drawn in this work are for cores that evolve in quasistatic manner so that they are probably only true for cores in relatively quiescent clouds where rapid increase in the magnitude of external pressure is unlikely.

\section{Conclusions}
In light of the results obtained from numerical simulations discussed in this work we argue, the fate of a prestellar core bears largely upon its temperature profile. Our realisation of the core  {\small B68} with different choices of the initial distribution of gas shows that the likely fate of a test core bears more strongly on the temperature profile that it develops over the course of its evolution. For even the test core {\small B68} modelled as a unstable {\small BES}, like other test cores that remained starless in this numerical exercise, rebounded soon after acquiring a centrally peaked density profile. Furthermore, other cores that remained starless were seen to mimic the profile of a unstable {\small BES} at the epoch when they were centrally most concentrated and yet, remained starless. We note, some of our test cores viz., {\small L694-2, L1689-SMM2} and {\small L1521F}(with standard dust grain-size) did in fact become thermally super-critical before rebounding, while the remaining two viz., {\small B68} and {\small L1517B} remained sub-critical with a peak density less than 10$^{5}$ cm$^{-3}$. It must be noted here that the core {\small L1521F} became super-critical and eventually a {\small VeLLO} only with a larger dust grain-size. In fact, with a larger grain-size all other starless cores also developed approximately uniform cold interiors before eventually collapsing. These observations from the present work lead to the conclusion that thermal super-criticality is probably only a necessary condition, though not sufficient, to ensure that a core becomes protostellar. \\ \\
We suggest, the strength of gas-dust coupling in a core likely holds the key to the eventual fate that befalls the respective core. In this case,  we demonstrate, gas temperature within a core is lowered more uniformly. 
With the choice of grain-size for dust typically found in the {\small ISM} we observed that the core {\small L1521F} developed a temperature profile similar to that observed in a typical starless core such as the {\small B68}. In this latter case gas-temperature in the core was approximately uniform, on the order of 12 K-14 K and showed a gradual reduction to $\lesssim$10 K only very close to its centre (see plots in Fig. 10). The conclusion about fluffy dust grains is consistent with the earlier suggestion by Keto \& Caselli (2008) in this regard. This idea is presently supported by at least tentative evidence from a few detailed studies of dust properties towards some prestellar cores and clumps. Our result here therefore calls for more studies about the nature of dust grains in prestellar cores and in cores that remain starless. \\ \\
Furthermore, realisations of cores that remained starless in this work also demonstrate that these cores can indeed survive over at least a few free-fall times (see Table 2). In fact realisations discussed here reproduce the typical core contraction time, the in-fall velocities as well as the rate of accretion for a typical {\small VeLLO}. These simulations also show that the virial state of a core changes over the course of its evolution. We tracked this change on the so-called virial chart that marks the temporal variation of virial components of a core.  Cores were seen to evolve from a state that was pressure-confined to one that was gravitationally bound. Similarly, we have also shown that cores make a transition from a state that is less virially bound to one that is more strongly virially bound.  Listed in Table 3 are the magnitudes of virial coefficients $Q$, and $X$, for respective cores modelled in this work. The magnitudes of these coefficients for cores that remained starless in this work call for attention; not only are these cores virially bound ($X >$ 1), but also relatively strongly bound ($Q >$1), except {\small L1517B}. This suggests, gravitational and/or virial boundedness of a core is not a sufficient condition for it to collapse, though it is probably necessary. A prospective protostellar core must continue to cool efficiently as it continues to contract in quasistatic manner. In the foreseeable future we propose to expand the remit of this work by coupling the hydrodynamic calculations discussed here by including the effects of depletion of various molecular species as a core continues to acquire higher densities during its collapse. Also, we should like to examine the impact of ambient environment on the evolution of cores and furthermore, include the magnetic field in these calculations to study how it affects the dynamic evolution of a core.

\begin{acknowledgements}
  The author gratefully acknowledges the warm hospitality extended by the IAAT, T$\ddot{\mathrm{u}}$bingen, where this work was carried out. Also special thanks to James Di Francesco and Derek Ward-Thompson for kindly proof-reading this text. Simulations discussed in this work were developed on the bwGRiD computing cluster, a member of the German D-Grid initiative, funded by the Ministry for Education and Research (Bundesministerium f$\ddot{\mathrm{u}}$r Bildung und Forschung) and the Ministry for Science, Research and Arts Baden-W$\ddot{\mathrm{u}}$rttemberg (Ministerium f$\ddot{\mathrm{u}}$r Wissenschaft, Forschung und Kunst Baden-W$\ddot{\mathrm{u}}$rttemberg). The author gratefully acknowledges useful inputs from Rolf Kuiper and Katherine Pattle. This work was partially supported by the Young Scientist Award (YSS/2014/000304) of the Department of Science \& Technology of the Government of India.  
\end{acknowledgements}



\begin{thebibliography}{}
 \bibitem[\protect\citeauthoryear{Aguti}{2007}]{b1}Aguti, E. D., Lada, C. J., Bergin, E., Alves, J \& Berkinshaw, M., 2007, ApJ, 665, 457
\bibitem[\protect\citeauthoryear{Alves}{2001}]{b2}Alves, J., Lada, C \& Lada, E., 2001, Nature, 409, 159
\bibitem[\protect\citeauthoryear{Anathpindika}{2011}]{b3}Anathpindika, S., 2011, New Astronomy, 16, 477
\bibitem[\protect\citeauthoryear{Anathpindika}{2013}]{b4}Anathpindika, S \& Di Francesco, 2013, MNRAS, 430, 1854
\bibitem[\protect\citeauthoryear{Anathpindika}{2015}]{b5} Anathpindika, S., 2015, PASA, 32, 2
\bibitem[\protect\citeauthoryear{Bailey}{2014}]{b6} Bailey, N., Banerjee, R \& Caselli, P., 2014, ApJ, 798, 75
\bibitem[\protect\citeauthoryear{Ballesteros}{2003}]{b7} Ballesteros-Paredes, J., Klessen, R \& V{\' a}zquez-Semadeni, E., 2003, ApJ, 592, 188 
\bibitem[\protect\citeauthoryear{Balesteros}{2006}]{b8} Ballesteros-Paredes, J., Gazol, A., Kim, J., Klessen, R., Jappsen, A.-K \& Jejero, E., 2006, ApJ, 667, 384 
\bibitem[\protect\citeauthoryear{Bate}{1995}]{b9}Bate, M. R., Bonnell, I. A. \& Price, N. M., 1995, MNRAS, 277, 362
\bibitem[\protect\citeauthoryear{Burkert}{1997}]{b10}Bate, M. R \& Burkert, A., 1997, MNRAS, 288, 1060
\bibitem[\protect\citeauthoryear{Bate2}{2015}]{b99}Bate, M \& Keto, E., MNRAS, 2015, 449, 2643
\bibitem[\protect\citeauthoryear{Beichman}{1989}]{b11}Beichman, C. A., Myers, P. C., Emerson, J. P., Harris, S., Mathieu, R.,
Benson, P. J., \& Jennings R. E. 1986, ApJ, 307, 337
\bibitem[\protect\citeauthoryear{Benson}{1989}]{b12}Benson, P. J \& Myers, P. C., 1989, ApJS, 71, 89
\bibitem[\protect\citeauthoryear{Bergin}{2006}]{b13} Bergin, E., Maret, S., Van der Tak, F. S., Alves, J., Cardmody, S \& Lada, C. J., ApJ, 2006, 645, 369
\bibitem[\protect\citeauthoryear{Bodenheimer}{1968}]{b14} Bodenheimer, P \& Sweigert, A., 1968, ApJ, 152, 515
\bibitem[\protect\citeauthoryear{Bondi}{1952}]{b15} Bondi, F., 1952, MNRAS, 112, 195
\bibitem[\protect\citeauthoryear{Bonnor}{1955}]{b16} Bonnor, W., 1956, MNRAS, 116, 351
\bibitem[\protect\citeauthoryear{Bourke}{1995}]{b17} Bourke, T., Hyland, A., Robinson, G., James, S \& Wright, C., 1995, MNRAS, 276, 1067
\bibitem[\protect\citeauthoryear{Bourke}{2006}]{b18} Bourke, T. L., Myers, P. C., Evans II, N. J.,  Dunhan, M. M., Kauffmann, J. \emph{et al.}, 2006, ApJ, 649, L37
\bibitem[\protect\citeauthoryear{Burke}{1983}]{b101}Burke, J. R \& Hollenbach, D., J., 1983, ApJ, 265, 223
\bibitem[\protect\citeauthoryear{Caselli}{2010}]{b20} Caselli, P., Keto, E., Pagani, L., Aikawa, Y \emph{et al.}, 2010, A\&A, 521, L29
\bibitem[\protect\citeauthoryear{Caselli}{2010}]{b21} Caselli, P., Walmsley, C. M., Zucconi, A., Tafalla, M., Dore, L \& Myers, P. C., 2002, ApJ, 565, 344
\bibitem[\protect\citeauthoryear{Chandrasekar}{1939}]{b22} Chandrasekar, S., 1939, \emph{An Introduction to the study of stellar structure}, Dover Pub. Inc.
\bibitem[\protect\citeauthoryear{Chen}{2015}]{b19}Chen Y-C, M., Di Francesco, J., Johnstone, D., Sadavoy, S., Hatchell, J., Mottram, H \emph{et al.}, 2015, \emph{ApJ submitted.}
\bibitem[\protect\citeauthoryear{Chitsazzadeh}{2014}]{b23}Chitsazzadeh, S., 2014, \emph{PhD. dissertation}, University of Victoria
\bibitem[\protect\citeauthoryear{Chitsazzadeh}{2014}]{b24}Chitsazzadeh, S., Di Francesco, J., Schnee, S., Friesen, R. K \emph{et al.}, 2014, ApJ, 790, 124
\bibitem[\protect\citeauthoryear{Ciolek}{2001}]{b25}Ciolek, G \& Basu, S., 2001, ApJ, 547, 272
\bibitem[\protect\citeauthoryear{Crapsi}{2004}]{b26}Crapsi, A., Caselli, P., Walmsley, C. M., Tafalla, M., Lee, C. W., Bourke, T. L. \& Myers, P. C., 2004, A\&A, 2004, 420, 957
\bibitem[\protect\citeauthoryear{Crapsi}{2007}]{b27}Crapsi, A., Caselli, P., Walmsley, M \& Tafalla, M., 2007, A\&A, 470, 221
\bibitem[\protect\citeauthoryear{Francesco}{2007}]{b28}Di Francesco, J., Evans II, N. J., Caselli, P., Myers, P. C., Shirley, Y., Aikawa, Y \& Tafalla, M., 2007, Protostars \& Planets V, Eds. B. Reipurth, D. Jewitt \& K. Keil. \emph{Tuscon AZ: Arizona University Press}, 17
\bibitem[\protect\citeauthoryear{Ebert}{1955}]{b29} Ebert, R., 1955, ZA, 36, 22
\bibitem[\protect\citeauthoryear{Elmegreen}{2000}]{b30} Elmegreen, B., ApJ, 530, 277
\bibitem[\protect\citeauthoryear{Enoch}{2007}]{b31} Enoch, M. L., Glenn, J., Evans, N. J., II, Sargent, A. I., Young, K. E., \& Huard, T. L. 2007, ApJ, 666, 982
\bibitem[\protect\citeauthoryear{Evans}{2001}]{b32} Evans, N. J II., Rawlings, J. M. C., Shirley, Y \& Mundy, L., 2001, ApJ, 557, 193
\bibitem[\protect\citeauthoryear{Federatth}{2010}]{b33} Federrath, C., Banerjee, R., Clark, P., C \& Klessen, R., 2010, ApJ, 713, 269
\bibitem[\protect\citeauthoryear{Federatth}{2014}]{b34} Federrath, C., Schr$\ddot{\mathrm{o}}$n, M., Banerjee, R \& Klessen, R., 2014, ApJ, 790, 128
\bibitem[\protect\citeauthoryear{Fu}{2011}]{b35} Fu, T. M., Gao, Y \& Lou, Y. Q., 2011, ApJ, 741, 113
\bibitem[\protect\citeauthoryear{Goldsmith}{2001}]{b97}Goldsmith, P., 2001, ApJ, 557, 736
\bibitem[\protect\citeauthoryear{gOMEZ}{2007}]{b36} G{\' o}mez, G., V{\' a}zquez-Semadeni, E., Shadmehri, M \& Ballesteros-Paredes, j., 2007, ApJ, 669, 1042
\bibitem[\protect\citeauthoryear{Gong}{2009}]{b37} Gong, H \& Ostriker, E., 2009, ApJ, 699, 230
\bibitem[\protect\citeauthoryear{Gong}{2011}]{b38} Gong, H \& Ostriker, E., 2011, ApJ, 729, 120
\bibitem[\protect\citeauthoryear{Gregersen}{2000}]{b39}Gregersen, E. M. \& Evans, N. J., 2000, ApJ, 538, 260
\bibitem[\protect\citeauthoryear{Harvey}{2003}]{b40}Hartmann, L., Ballesteros-Paredes, J \& Bergin, E. A., 2001, ApJ, 562, 852
\bibitem[\protect\citeauthoryear{Harvey}{2003}]{b41}Harvey, D. W. A., Wilner, D. J., Lada, C., Myers, P. C., Alves, J., 2003, ApJ, 598, 1112
\bibitem[\protect\citeauthoryear{Hennebelle}{2003}]{b42}Hennebelle, P., Whitworth, A. P., Gladwin, P. P \& Andr{\' e}, Ph., 2003, MNRAS, 340, 870
\bibitem[\protect\citeauthoryear{Henriksen}{1997}]{b43}Henriksen, R., Andr{\' e} , Ph., \& Bontemps, S., 1997, A\&A, 323, 549
\bibitem[\protect\citeauthoryear{Hubber et al.}{2006}]{b44} Hubber, D., Goodwin, S. \& Whitworth, A. P., 2006, A\&A, 450, 881
\bibitem[\protect\citeauthoryear{Hubber}{2011}]{b45}Hubber, D., Batty, C., McLeod, A \& Whitworth, A. P., 2011, A\&A, 529, 28
\bibitem[\protect\citeauthoryear{Johnstone}{2000}]{b46}Johnstone, D., Wilson, C. D., Christine, D., Moriarty-Schieven, G., Joncas, G., Smith, G., Gregersen, E \& Fich, M., 2000, ApJ, 545, 327
\bibitem[\protect\citeauthoryear{Juvela}{2015}]{b112}Juvela, M., Demyk, K., Doi, Y., Hughes, A. \emph{et al.}, 2015, astroph. 1509.08023
\bibitem[\protect\citeauthoryear{Kaminski}{2014}]{b47}Kaminski, E., Frank, A., Jonathan, S \& Myers, P. C., 2014, ApJ, 790, 70
\bibitem[\protect\citeauthoryear{Kauffmann}{2005}]{b48}Kauffmann,  J., Bertoldi, F. Evans II, N. J., C2D Collaboration, 2005, Astronomische Nachrichten, 326, 378
\bibitem[\protect\citeauthoryear{Field}{2004}]{b107}Keto, E., Rybicki, G., Bergin, E \& Plume, R., 2004, ApJ, 613, 355
\bibitem[\protect\citeauthoryear{Field}{2005}]{b49}Keto, E \& Field, G., 2005, ApJ, 635, 1151
\bibitem[\protect\citeauthoryear{Brod}{2006}]{b50}Keto, E., Broderick, E., Lada, C \& Narayan, R., 2006, ApJ, 652, 1366
\bibitem[\protect\citeauthoryear{Keto}{2008}]{b51}Keto, E \& Caselli, P., 2008, ApJ, 683, 238
\bibitem[\protect\citeauthoryear{Keto}{2010}]{b52}Keto, E \& Caselli, P., 2010, MNRAS, 402, 1625
\bibitem[\protect\citeauthoryear{Keto}{2015}]{b53}Keto, E., Caselli, P \& Rawlings, J. M. C., 2015, MNRAS, 446, 3731
\bibitem[\protect\citeauthoryear{Kirk}{2005}]{b54}Kirk, J. M., Ward-Thompson, D \& Andr{\' e}, Ph., 2005, MNRAS, 360, 1506
\bibitem[\protect\citeauthoryear{Kirk}{2006}]{b55}Kirk, J. M., Ward-Thompson, D \& Crutcher, R., 2006, MNRAS, 369, 1445
\bibitem[\protect\citeauthoryear{Kruegel}{1994}]{b105}Kr$\ddot{u}$gel, E \& Siebenmorgen, R., 1994, A\&A, 288, 929
\bibitem[\protect\citeauthoryear{Lada}{2003}]{b56}Lada, C., Bergin, E. A., Alves, J. F \& Huard, T. L., 2003, ApJ, 586, 286
\bibitem[\protect\citeauthoryear{Larson}{1969}]{b57}Larson, R., 1969, MNRAS, 145, 271
\bibitem[\protect\citeauthoryear{Lee}{1999}]{b58}Lee, C. W. \& Myers, P. C., 1999, ApJS, 123, 233
\bibitem[\protect\citeauthoryear{Lee}{1999}]{b59}Lee, C. W. \& Myers, P. C \& Tafalla, M., 1999, ApJ, 526, 788
\bibitem[\protect\citeauthoryear{Lee}{2004}]{b60}Lee, C. W., Myers, P. C \& Plume, R., 2004, ApJS, 153, 523 
\bibitem[\protect\citeauthoryear{Lee}{2007}]{b61}Lee, S. H., Park, Y.-S., Sohn, J., Lee, S. H \& Lee, H. M., 2007, ApJ, 660, 1326
\bibitem[\protect\citeauthoryear{Levshakov}{2013}]{b62}Levshakov, S., Henkel, C., Reimers, D., Wang, M., Mao, R., Wang, H \& Xu, Y., 2013, A\&A, 553, A58
\bibitem[\protect\citeauthoryear{Mac}{1998}]{b63}Mac Low, M. M., Klessen, R., Burkert, A \& Smith, M., 1998, PhRvL, 80, 2754
\bibitem[\protect\citeauthoryear{Maret}{2011}]{b64}Maret, S., Bergin, E \& Lada, C., 2007, ApJ, 670, L25
\bibitem[\protect\citeauthoryear{Maret}{2013}]{b65}Maret, S., Bergin, E \& Tafalla, M., 2013, A\&A, 559, 53
\bibitem[\protect\citeauthoryear{Monaghan}{1997}]{b66} Monaghan, J., 1997, JCoPh, 136, 298
\bibitem[\protect\citeauthoryear{Mouschovias}{1991}]{b67} Mouschovias, T., 1991, ApJ, 373, 169
\bibitem[\protect\citeauthoryear{nAKANO}{1998}]{b68}Nakano, T., 1998, ApJ, 494, 587
\bibitem[\protect\citeauthoryear{nIEL}{2012}]{b69}Nielbock, M., Launhardt, R., Steinacker, J., Stutz, A. M., Balog, Z \emph{et al.}, 2012, A\&A, 547, 11
\bibitem[\protect\citeauthoryear{Offner}{2014}]{b100}Offner, S., Clark, P., Hennebelle, P., Bastian, N., Bate, M., Hoppkins, P., Moraux, E \& Whitworth, A., 2014, Protostars \& Planets VI, 53, University of Arizona Press, Eds. Beuther, H., Klessen, R, Dullemond, C \& Henning, Th.
\bibitem[\protect\citeauthoryear{Onishi}{1999}]{b70}Onishi, T., Mizuno, A \& Fukui, Y., 1999, PASJ, 51, 257
\bibitem[\protect\citeauthoryear{Ossenlopf}{1994}]{b106}Ossenkopf, V \& Henning, Th., 1994, A\&A, 291, 943
\bibitem[\protect\citeauthoryear{PattLe}{2015}]{b71}Pattle, K., Ward-Thompson, D., Kirk, J. M., White, G. J., Drabek-Maunder, E., Buckle, J \emph{et al.}, 2015, MNRAS, 450, 1094
\bibitem[\protect\citeauthoryear{Penston}{1969}]{b72}Penston, M. V., 1969, MNRAS, 144, 425
\bibitem[\protect\citeauthoryear{Redman}{2004}]{b143}Redman, M. P., Rawlings, J. M. C., Yates, J \& Williams, D. A., 2004, MNRAS, 352, 243 
\bibitem[\protect\citeauthoryear{Redman}{2006}]{b73}Redman, M. P., Keto, E \& Rawlings, J. M. C., 2006, MNRAS, 370, L1 
\bibitem[\protect\citeauthoryear{Roy}{2014}]{b74}Roy, A., Andr{\' e}, Ph, Palmeirim, P., Attard, M., K$\ddot{\mathrm{o}}$nyves, V \emph{et al.}, 2014, MNRAS, A\&A, 562, 138
\bibitem[\protect\citeauthoryear{Sadavoy}{2010a}]{b75}Sadavoy, S., Di Francesco, J., Bontemps, S., Megeath, S., Rebull, S \emph{et al.}, 2010a, ApJ, 710, 1247
\bibitem[\protect\citeauthoryear{Sadavoy}{2010b}]{b76}Sadavoy, S., Di Francesco, J \& Johnstone, D., 2010b, ApJ, 718, L32
\bibitem[\protect\citeauthoryear{Safier}{1997}]{b77}Safier, P. N., McKee, C. F., \& Stahler, S. W., 1997, ApJ, 485, 660
\bibitem[\protect\citeauthoryear{Schnee}{2013}]{b78}Schnee, S., Brunetti, N., Di Francesco, J., Caselli, P., Friesen, R., Johnstone, D \& Pon, A., 2013, ApJ, 777, 121
\bibitem[\protect\citeauthoryear{Shu}{1977}]{b79}Shu, F., 1977, ApJ, 214, 488
\bibitem[\protect\citeauthoryear{Shu}{1987}]{b80}Shu, F., Adams, F. C \& LIzano, S., 1987, Ann. Rev. A\&A, 25, 23
\bibitem[\protect\citeauthoryear{Tafalla}{1998}]{b82}Tafalla, M., Mardones, D., Myers, P. C., Caselli, P., Bachiller, R \& Benson, P. J., 1998, ApJ, 504, 90
\bibitem[\protect\citeauthoryear{Tafalla}{2002}]{b83}Tafalla, M., Myers, P. C., Caselli, P., Walmsley, C. M \& Comito, C., 2002, ApJ, 569, 815
\bibitem[\protect\citeauthoryear{Tafalla}{2004}]{b84}Tafalla, M., Myers, P. C., Caselli, P \& Walmsley, C. M., 2004, A\&A, 416, 191
\bibitem[\protect\citeauthoryear{Takahashi}{2013}]{b85}Takahashi, S., Ohashi, N \& Bourke, T. L., 2013, ApJ, 774, 20
\bibitem[\protect\citeauthoryear{Teixeira}{2005}]{b86}Teixeira, P., Lada, C \& Alves, J., 2005, ApJ, 629, 726
\bibitem[\protect\citeauthoryear{Terebey}{2009}]{b87}Terebey, S., Fich, M., Noriega-Crespo, A., Padgett, D. L., Fukagawa, M., Audard, M \emph{et al.}, 2009, ApJ, 696, 1918
\bibitem[\protect\citeauthoryear{Testi}{2014}]{b111}Testi, L., Birnstiel, T., Ricci, L., Andrews, S \emph{et al.}, 2014, 339, PPVI, \emph{Arizona University Press}, Tuscon, Eds. Beuther, H., Klessen, R., Dullemond, C \& Henning, Th.
\bibitem[\protect\citeauthoryear{Truelove}{1997}]{b88}Truelove, J. K., Klein, R. I., McKee, C. F., Holliman, J. H., Howell, L. K \& Greenhough, J. A., 1997, ApJ, 489, L179 
\bibitem[\protect\citeauthoryear{Valdivia}{2016}]{b150}Valdivia, V., Hennebelle, P., G{\' e}rin, M \& Lesaffre, P., 2016, astroph 1512.05523, \emph{To appear in A\&A}
\bibitem[\protect\citeauthoryear{Walch}{2010}]{b89}Walch, S., Naab, T., Whitworth, A., Burkert, A \& Gritschneder, M., 2010, MNRAS, 402, 2253 
\bibitem[\protect\citeauthoryear{Ward-thompson}{1994}]{b90}Ward-Thompson, D., Scott, P. F., Hills, R. E., \& Andr{\' e} , Ph. 1994, MNRAS, 268, 276
\bibitem[\protect\citeauthoryear{Ward-thompson}{2007}]{b91}Ward-Thompson, D., Andr{\' e}, P., Crutcher, R., Johnstone, D., Onishi, T., \& Wilson, C. 2007, in Protostars and Planets V, ed. B. Reipurth, D. Jewitt, \& K. Keil (Tucson: Univ. Arizona Press), 33
\bibitem[\protect\citeauthoryear{Whitworth}{1985}]{b92}Whitworth, A \& Summers, D., 1985, 214, 1
\bibitem[\protect\citeauthoryear{Whitworth}{2001}]{b93}Whitworth, A \& Ward-Thompson, D., 2001, ApJ, 547, 317
\bibitem[\protect\citeauthoryear{Williams}{1999}]{b94}Williams, J. P., Myers, P. C., Wilner, D \& Di Francesco, J., 1999, ApJ, 513, L61
\bibitem[\protect\citeauthoryear{Williams}{2006}]{b95}Williams, J. P., Lee, C. W \& Myers, P. C., 2006, ApJ, 636, 952
\bibitem[\protect\citeauthoryear{Young}{2004}]{b96}Young, C. H., J$\oslash$rgensen, J., Shirley, Y., Kauffmann, J., Huard, T., Lai, S \emph{et al.}, 2004, ApJ, 154, 396 
\end{thebibliography}

\end{document}